\documentclass{aa}
\usepackage{graphicx, graphics}
\usepackage{CJK}
\usepackage{bbm}
\usepackage{amsmath, amssymb, bm}
\usepackage[nodayofweek]{datetime}
\input{prepcitationformat.txt}
\usepackage{txfonts}
\usepackage{ulem}
\usepackage{booktabs}  %  引入三线表宏包
\usepackage[switch]{lineno}
\usepackage{tikz}

\definecolor{lime}{HTML}{A6CE39}
\DeclareRobustCommand{\orcidicon}{%
    \raisebox{-3pt}{\begin{tikzpicture}
    \filldraw [lime, yshift=-2pt] (0, 0) circle [radius=0.16]
    node[white] {\raisebox{1pt}{\hspace{0.5pt}\fontfamily{qag}\selectfont\tiny i\scalebox{0.8}{D}}};
    \end{tikzpicture}}
    \hspace{-2.5mm}
    \vspace{-0.25pt}
}

\global\def\tablenotemark#1{{\color{blue}{\normalfont\textsuperscript{\scriptsize #1}}}} % changed by Bin Ren to \scriptsize to match published tablenotemark

\newdateformat{monthyearday}{%
  \THEYEAR\ \monthname[\THEMONTH] \twodigit{\THEDAY}}
\newdateformat{daymonthyear}{%
  \twodigit{\THEDAY}\ \monthname[\THEMONTH] \THEYEAR}

\newcommand{\orcidauthor}[2]{#2\href{http://orcid.org/#1}{\orcidicon}}

\titlerunning{J1604 reflectance \& shadow}
\authorrunning{Zhong et al.}

\begin{document}
\begin{CJK*}{UTF8}{gbsn}
\title{Multi-band reflectance and shadowing of RX~J1604.3-2130 protoplanetary disk in scattered light\thanks{FITS images for Fig.~\ref{fig:fig-all} are only available at the CDS via anonymous ftp to \url{cdsarc.cds.unistra.fr} (\url{130.79.128.5}) or via \url{https://cdsarc.cds.unistra.fr/viz-bin/cat/J/A+A/}}}

\author{
\orcidauthor{0009-0005-9413-9840}{Huisheng Zhong (钟惠生)}\inst{\ref{inst-sysu}, \ref{inst-sysu-csst}}
\and
\orcidauthor{0000-0003-1698-9696}{Bin B. Ren (任彬)\thanks{Marie Sk\l odowska-Curie Fellow}}\inst{\ref{inst-oca}, \ref{inst-uga}, \ref{inst-cit}} %$^,$\thanks{To whom correspondence should be addressed.} <- Commented out since A&A told me they don't do it.
\and
\orcidauthor{0000-0002-0378-2023}{Bo Ma (马波)}\inst{\ref{inst-sysu}, \ref{inst-sysu-csst}} %$^{\star\star}$
\and
\orcidauthor{0000-0002-6318-0104}{Chen Xie (谢晨)}\inst{\ref{inst-jhu}, \ref{inst-lam}} %H1 H2 K1 K2
\and
\orcidauthor{0000-0003-3583-6652}{Jie Ma (马颉)}\inst{\ref{inst-ethz}} %J
\and
\orcidauthor{0000-0003-0354-0187}{Nicole L. Wallack} \inst{\ref{inst-dtm}, \ref{inst-gps}}
\and
\orcidauthor{0000-0002-8895-4735}{Dimitri Mawet} \inst{\ref{inst-cit}, \ref{inst-jpl}}
\and
\orcidauthor{0000-0003-4769-1665}{Garreth Ruane} \inst{\ref{inst-jpl}}
% \and 
% Others (mainly Caltech proposal team) % to be specified
}

\institute{
School of Physics and Astronomy, Sun Yat-sen University, Zhuhai, Guangdong 519082, China; \url{mabo8@mail.sysu.edu.cn} \label{inst-sysu}
\and
Center of CSST in the great bay area, Sun Yat-sen University, Zhuhai, Guangdong 519082, China \label{inst-sysu-csst}
\and
Universit\'{e} C\^{o}te d'Azur, Observatoire de la C\^{o}te d'Azur, CNRS, Laboratoire Lagrange, Bd de l'Observatoire, CS 34229, 06304 Nice cedex 4, France; \email{\url{bin.ren@oca.eu}} \label{inst-oca}
\and
Universit\'{e} Grenoble Alpes, CNRS, Institut de Plan\'{e}tologie et d'Astrophysique (IPAG), F-38000 Grenoble, France \label{inst-uga}
\and
Department of Astronomy, California Institute of Technology, MC 249-17, 1200 E California Blvd, Pasadena, CA 91125, USA \label{inst-cit}
\and
Department of Physics and Astronomy, The Johns Hopkins University, 3701 San Martin Drive, Baltimore, MD 21218, USA \label{inst-jhu}
\and
Aix Marseille Univ., CNRS, CNES, LAM, Marseille, France \label{inst-lam}
\and
Institute for Particle Physics and Astrophysics, ETH Z\"{u}rich, Wolfgang Pauli Strasse 17, 8093 Zurich, Switzerland\label{inst-ethz}
\and
Earth and Planets Laboratory, Carnegie Institution for Science, Washington, DC 20015, USA \label{inst-dtm}
\and
Division of Geological \& Planetary Sciences, California Institute of Technology, MC 150-21, Pasadena, CA 91125, USA \label{inst-gps}
\and
Jet Propulsion Laboratory, California Institute of Technology, 4800 Oak Grove Dr, Pasadena, CA, 91109, USA \label{inst-jpl}
}

\date{Received 07 December 2023; revised 01 February 2024; accepted 08 February 2024}

\abstract
{Spatially-resoved cicrumstellar disk spectrum and composition can provide valuable insights into the bulk composition of forming planets, as well as the mineralogical signatures that emerge during and after planet formation.}
{We aim to systemically extract the RX~J1604.3-213010 (J1604 hereafter) protoplanetary disk in high-contrast imaging observations, and obtain its multi-band reflectance in visible to near-infrared wavelengths.}
{We obtained coronagraphic observations of J1604 from the Keck Observatory's NIRC2 instrument, and archival data from the Very Large Telescope's SPHERE instrument. Using archival images to remove star light and speckles, we recovered the J1604 disk and obtained its surface brightness using forward modeling. Together with polarization data, we obtained the relative reflectance of the disk in $R$, $J$, $H$ ($H2$ and $H3$), $K$ ($K1$ and $K2$), and $L'$ bands spanning two years.}
%{We retrieved  archival data of J1604 from the ESO archive and resolved the images of the disk by employing the KLIP method. Subsequently, We derived the optimal models for the disk in multiple wavelengths with forward modeling. And we measured the host star flux  of J1604  with aperture photometry. Additionally, We measured the flux of the disk using the models or the observation with throughput correction. To enhance the dataset, we measured the flux of the disk and the star in the $R$, $J$ and $H$ bands using the data and the models inferred by \citet{2023A&AMa}.  Consequently, we obtained the reflectance spectrum of J1604 and compared it with other disk.}
{Relative to the J1604 star, the resolved disk has a reflectance of ${\sim}10^{-1}$~arcsec$^{-2}$ in $R$ through $H$ bands and ${\sim}10^{-2}$~arcsec$^{-2}$ in $K$ and $L'$ bands, showing a blue color. Together with other systems, we summarized the multi-band reflectance for 9 systems. We also identified varying disk geometry structure, and a shadow that vanished between June and August in 2015.}
{Motivated by broad-band observations, the deployment of cutting-edge technologies could yield higher-resolution reflection spectra, thereby informing the dust composition of disks in scattered light in the future. With multi-epoch observations, variable shadows have the potential to deepen insights into the dynamic characteristics of inner disk regions.}
%{We detected  dynamic shadows and one potential  inner dust structures and the resulting spectrum revealed a pronounced decreasing trend with increasing wavelength.}
%{Comparing chemical abundances of a planet and the host star reveals the origin and formation pathway of the planet. }


\maketitle

\section{Introduction}

% General layout for introduction: 

% 1. Status of understanding of composition for planets (atmosphere, bulk composition)

% 2. Existing inference on bulk composition

% 3. Advantages and limitations of existing inferences

% 4. What can we do to move towards solving the problem

% 4.1 Existing studies on other systems (photometry, reflectance spectroscopy, transmission spectrum, emission spectrum);

% 4.2 What does the theorists say on composition

% 4.3 Existing studies on J1604, and now with Keck and SPHERE, we can extend to longer wavelengths, etc.\\

% Section 3. Discussion:

% 1. Compare with HR~4796A and HD~32297 --- debris disk (giant planet formation is done). J1604: protoplanetary disk (giant planet formation ongoing)

% 2. Limitation: light-travel time (\url{https://arxiv.org/pdf/2304.07370.pdf}) also search "light echo". We can also include TESS light curves. \\

Over 5000 exoplanets have been found  with different observational techniques to date,\footnote{\url{https://exoplanetarchive.ipac.caltech.edu/}} and the diversity in their size and mass distribution demonstrates the variety of the formation and evolution processes of planetary systems. Planets are formed within circumstellar disks around stars, implying that all are made from gas and dust inherited from the same molecular cloud. They could thus share similar bulk composition \citep[e.g.,][]{2020wangji}, suggesting that the composition of planets, disks, and stars are correlated. However, planets can form from different mechanisms, primarily through core accretion \citep[e.g.,][]{Pollack1996} and  disk gravitational instability \citep[e.g.,][]{Pollack1996, Piso2014, Piso2015}, and these models predict different planetary luminosity and spectra \citep[e.g.,][]{Spiegel2012ApJ}. An investigation into the compositional makeup of planetary systems -- including planets and disks -- can contribute to our understanding of these celestial bodies, offering an opportunity to empirically test prevailing theories of planet formation.

%Thus, probing the planetary composition is an effective way to constrain the planet formation mechanism.

Various indirect techniques were proposed to infer the bulk composition of exoplanets. Typically, mass-radius measurements of exoplanets are employed to estimate the planetary bulk composition \citep[e.g.,][]{Zeng2019,Miller2011ApJ,Thorngren2016,ller2020ApJ,Plotnykov2020MNRAS,Adibekyan2021}. %\sout{Studies using different evolution models to infer the metallicity of warm giant planets with  moderate stellar insolations fluxes  found a strong relation between  the metal-enrichment, signifying the ratio of the metallicity between the planet and the host star, and the planetary mass Miller2011ApJ,Thorngren2016,ller2020ApJ.Additionally, Plotnykov2020MNRASfound that the bulk composition of rocky planets spans a wider range than stars. Similarly, when estimating the iron-mass fraction from the masses and radii of the rocky exoplanets, citet{Adibekyan2021} found that planetary iron mass fraction correlates with host star iron mass fraction.  Planets with small radii ($2\sim4R_{\oplus}$) require a gas envelope of at most a few mass fraction, and their masses are dominated by their cores. citet{Zeng2019} calculated the growth curves of planets which indicated the relationship between the mass and the radii in the process of adding either ice or gas to a rocky core. Using Monte Carlo simulation with the assumption that the ratio of rocky cores to icy cores is 1:2, citet{Zeng2019} reproduced successfully the bimodal radius distribution of the small exoplanets. }
%Using transmission spectra,  \citet{Pinhas2019} found that the majority of hot Jupiters have atmospheres consistent with sub-solar $H_{2}O$ abundances at their day-night terminators. \citet{Hu_2012} develop a theoretical framework to investigate reflection and thermal emission spectra of airless rocky exoplanets, proposing the potential spectroscopic detectability of their rocky surfaces through distinctive spectral features. Additionally, an asteroid whose orbit is perturbed, would pass within the white dwarf's tidal radius and be destroyed, and then the bodies would produce a circumstellar disk. Eventually this material is accreted onto the host star  \citep{Debes2002ApJ, Jura2003ApJ}. So spectroscopic determination of the abundances in externally polluted white dwarfs could reveal the elemental composition of the accreted planetary bodies \citep{Jura2014AREPS}. Similarly the photospheric abundances of the stars, which have radiative envelopes and hence less bulk–photosphere mixing, may  reveal the composition of the  material which were accreted from the circumstellar disks onto the host stars \citep{Jermyn2018MNRAS}. \citet{Kama2019ApJ} found $(89 \pm 8)\%$ of elemental sulfur is in refractory form in their disks by the use of accretion contamination on the surfaces of early-type stars. \citet{McClure2020A&A} measured the disk abundances inside the dust sublimation radius from near-infrared atomic emission lines.
%For small planets, mass measured using RV method,
%However, 
The uncertainties in exoplanet property measurements, including radius and mass are however large \citep[e.g.,][]{Weiss2014ApJ}. In addition, the uncertainties or degeneracy from theory predictions are significant \citep{ller2020ApJ, ller2023FrASS, Rogers2010ApJ, Dressing2015ApJ}. Consequently, they can not precisely constrain the bulk composition of exoplanets. 
%\sout{What is more, there is significant  theoretical uncertainties that can result into different exoplanet properties citep{ller2020ApJ, ller2023FrASS}: the composition of exoplanets is plagued with degeneracy Rogers2010ApJ, Dressing2015ApJ They cannot inform to what degree did physical processes, during nebular disk accretion versus post-nebular disk accretion (e.g., impact erosion), influence planets' final bulk composition citep{McDonough2020arXiv200904311M, McDonough2021PEPS}}. 
%The way protoplanetary disks evolve  impacts  the surface density distribution, $\Sigma (R)$, and the disk outer radius, $R_{out}$, two parameters that determine the planetesimal growth  and the planets migration, consequently the  architecture of the forming planetary systems \citep{Morbidelli2016Challenges}.  

Planet migration plays a decisive role in the evolution of rings and planets: the remnant planetesimal belts are candidates for cold debris discs \citep[e.g.,][]{Jiang2023}. %Scattered or shepherded by the formed planet, the radial profile of the planetesimal belt shows an asymmetry with a steep rise closer at its interior side and a more gradual decline towards the exterior regions. 
\citet{Morbidelli2016} showed that accretionary processes play a major role in determining a planet's bulk composition and volatile budget. %\citet{Jorge2021} showed that the composition of planets crucially depends on the abundances of the stellar system under investigation. 
%zengli 
%Direct iamged planet like HR~8799~c, Wang et al. (2020).
%行星的组成很重要，但是研究盘的组成也很重要，因为通过比较两者，可以让我们更深刻的理解行星的形成过程。
%"Conducting" or "Extracting" or "Inferring"
%\textcolor{red}{%Considering one system including both an exoplanet and a disk, 
%With the information about the grain composition of the disk and a speculative hypothesis about migration and precise numerical simulations of accretion, we could constrain material accreted in recent years and then provide the lower limit mass fraction of the exoplanet about the material that still remains in the disk.
In the case of young planets still embedded in protoplanetary disks, it thus might be possible to probe the planetary composition by obtaining the dust composition of the disks with a speculative hypothesis about migration and precise numerical simulations of accretion \citep{Pacetti2022ApJ,Mah2023A&A}. %}
Therefore, extracting disk composition might %\sout{reveal} 
contribute to the confirmation of the %\sout{bulk} 
composition of exoplanets. Characterizing the initial elemental budget contained in the protoplanetary disks from which giant planets are born could thus provide constraints to inform planet formation models \citep{Turrini2021ApJ,Pacetti2022ApJ}.
%Using disk composition, we can additionally relate to the composition of planets, as well as the compositional signatures during/after planet formation process.
%\sout{In the case of young planets still embedded in protoplanetary disks, it thus might be possible to directly access this information by probing the dust composition of the disks.}

Existing studies have attempted inferring the potential composition of circumstellar disks (e.g., protoplanetary, debris).
%HR~4796A is a young (8 Myr) star located 72.8 pc away from Earth \citep{Leeuwen2007} with a circumstellar debris disk.
%\citet{Debes2008} modeled the dust scatting efficiency of the disk at different wavelengths, and 
With broadband photometry, \citet{Debes2008} found that the HR~4796 debris ring might be explained using $1.4~\mu{\rm m}$-radius grains of complex organic  material;
%\citet{Rodigas2015} presented high-resolution images of the HR~4796A debris disk and 
\citet{Rodigas2015} found that silicates and organics are more generally favored over water ice for HR~4796, which showed possible existence of common constituents of both interstellar and solar system comets.
%\citet{Milli_2017} determine empirically the scattering phase function of the dust in the $H$ band and propose that the dust population is dominated by particles  of about 20$~\mum$ by analysing  the phase function, especially below 45$^\circ$. \citet{Milli2019} measured the morphology and scattering properties of the dust from the debris disk in polarised optical light over a wide range of scattering angles in the optical which is incompatible with dust particles being compact spheres under the assumption of the Mie theory.  \citet{Chen2020} used Gemini Planet Imager (GPI) multiband observations of HR~4796A to  measure the geometric parameter and the scattering phase function for the disk and then modeled the scattered phase function of the SPHERE $H2$ band using a distribution of hollow spheres composed of silicates, carbon and metallic iron.
Using integral field spectroscopy, \citet{Bhowmik2019} observed HD~32297 with Spectro-Polarimetric High-contrast Exoplanet REsearch \citep[SPHERE;][]{Beuzit2019} in $Y$, $J$, and $H$ bands in total intensity and found that the spectral reflectance of the debris disk features a ``gray to blue'' color, and interpreted that it resulted from the presence of grains far below the blowout size. %\textcolor{red}{
With a dip observed at the ice feature around $3.1~\mu$m, certain young disks showed potential presence of water ice, including HD~142527 %($\sim1$ Myr) 
\citep{Honda2009ApJ}, 
HD~100546 %($\lesssim 1 $ Myr) 
\citep{Honda2016ApJ}, AB~Aur %($4\pm1$ Myr) 
\citep{Betti2022AJ}, and HD 141569 %($5\pm3$ Myr) 
\citep{Kueny2024}. For HD~141569, \cite{Singh2021A&A} found a mild negative slope  and a absorption feature at around $1.5~\mu$m, potentially caused by the OH bonding resonance, in the reflectance across $Y$--$K_2$ bands. %} 
For complex circumstellar structures (e.g., spirals), a proper extraction of their morphology and reflectance is still challenging \citep[e.g.,][]{Olofsson2023A&A, Ren2023A&A}, since most existing algorithms suffer from overfitting or self-subtraction, which limit a proper recovery of disk signals in high-contrast imaging observations.  
%\citet{Olofsson2023A&A} found that DI-sNMF reduction \citep{Ren2020ApJ} compared to PCA reduction better recovered the flux of the projected semi-minor axis and the arcs, the birth ring. \citet{ Maire2017A&A}  obtained the reflectance spectra of different disk structures from the deprojected and $r^{2}$-scaled RDI images, which was  compared differently between different disk structures and demonstrated the different composition or sizes of the dust grains in the different disk structures. 

To properly recover disk images to study their reflectance in an early stage of disk evolution, here we study the RX~J1604.3-2130A (J1604 hereafter) disk in multiple wavelengths in scattered light with forward modeling. J1604 is a  $K2\pm1 $ star \citep{Preibisch2005} located at a distance of %$145.7\pm0.4$ pc with a mass of 1.24 $M_{\odot}$ \citep{Gaia2023A&A, Manara2020} 
$145.3\pm0.6$ pc \citep{Gaia2023A&A} with a mass of $1.46^{+0.19}_{-0.35}M_{\odot}$ \citep{Fouesneau2022A&A} and an age of $11\pm3$ Myr \citep{Pecaut2012}. \citet{Woitke2019PASP} modeled the spectral energy distribution (SED) of J1604, showing an infrared (IR) excess of 0.18$L_{\odot}$ for $\lambda > 6.72~\mu{\rm m}$ for a stellar luminosity of $L_{\star} = 0.76 L_{\odot}$. The transition outer disk around the star is  nearly face-on \citep[$6^{\circ}$ inclination,][]{Dong2017},  and massive with a dust mass of $\sim$40--50 $M_{\oplus}$ \citep{Barenfeld2016, Pinilla_2018}. There is evidence of planet-induced dust filtration \citep{Rice2006, Canovas2017}. \citet{hler2000} found that J1604 has a stellar companion located at $\sim$2300 au, itself being a binary with a 13~au separation. 
%A possible wide companion to J1604 at 16.22'' was identified by \citet{Kraus2009}. \citet{Kraus2009} cite Kunkel (1999),  K¨ohler et al. (2000)
\citet{Davies2019} calculated the misaligned angle between the stars' rotation axis and the outer regions of this transitional disk ($|i_{*}-i_{\rm disk} |>52^{\circ}$) at the $1.6 \sigma$ level. In addition, new measurements of the projected rotational velocity ($v \sin i $) indicated that the star is aligned with the inner disk, and thus misaligned with the outer disk \citep{Sicilia2020}.  

The outer disk has a deep resolved gas cavity that is smaller than the dust cavity \citep{Marel2015}. The  disk of J1604 was resolved with the Atacama Large Millimeter/submillimeter Array \citep[ALMA;][]{Mayama2018}, in which the observations are indicative of a misaligned inner disk with respect to the outer disk.  Considering the high-resolution continuum study of inner disks using ALMA, \citet{Francis2020} measured the dust mass, which gave an upper limit of 0.013 $M_{\oplus}$ for the inner disk. \citet{Stadler2023} modeled $^{12}$CO  intensity channel maps of the disk around J1604 and then obtained the position angle of the semimajor axis of the disk on the redshifted side ($\theta_{\rm PA} = 258\fdg7$), and suggested that another massive companion -- presumably orbiting with a significant inclination -- shapes the inner region spanning ${\sim}0\farcs25$ (${\sim}35$ au) based on localized non-Keplerian feature.

%\textcolor{red}{
In circumstellar disk systems, the misalignment between the inner disk and the outer disk could cast shadows on the outer disk, see \citet{Bohn2022A&A} for observation examples. Multi-epoch spectroscopic and near-infrared photometric observations reveal variability over several months, potentially linked to the instabilities or the perturbations within the inner disk \citep{Sitko2012ApJ}.  
In particular, changes in the shape, location and brightness of the shadow features provide valuable insights into the structure, and variability timescale of the dust casting shadow. For SAO~206462, \citet{Stolker2016A&A} witnessed varying shadow features, which maybe caused by a local perturbation of the inner disk or an accretion funnel
flow from the inner disk onto the star.  
Existing high-contrast imaging observations by \citet{Pinilla2018} found that the photometric %\sout{dips} \textcolor{red}{
shadows of J1604 outer disk are variable both in morphology and in location, suggesting that innermost regions are highly dynamic and thought could be evidence of a closer-in massive companion or a complex magnetic field topology. \citet{Ruane2019} observed similar features in $L'$ band with Keck. 
%The detected shadow features and their possible variability have the potential to provide insight into the structure of and processes occurring in the innermost disk regions.

In comparison with $J$-band data in \citet{Pinilla2018}, the scatter surface appears slightly further from the star in $L'$ band \citep{Ruane2019}, indicating the spatial segregation of dust particles sizes and lower opacities at longer wavelengths. \citet{Ma2022A&A} conducted precise measurements of intrinsic radiation parameters, including fractional polarization and apparent disk albedo, and comprehensively characterized the scattering behavior of dust within the disk utilizing a transition disk model. Therefore, obtaining multi-band reflectance for J1604 presents an opportunity to enhance our capacity to probe the bulk composition of this disk, as well as confirming the variability of the shadows on the outer disk to probe the inner disk. %In this study we obtained the relative reflectance of J1604 across multiple wavelengths. 
In Sect.~\ref{sec-obs}, we describe the details of the J1604 observations used in this study, and our data reduction procedure. Sect.~\ref{sec-ana}  encompasses our modeling result and the reflectance derived from the observations. We explore the substructure of J1604 and discuss the limitation of the method within Sect.~\ref{sec-dis}, which also features a comparative analysis of the relative reflectance between J1604 and other disks.  Finally, we provide a concise summary in Sect.~\ref{sec-sum}.

\section{Observation \& Data Reduction}\label{sec-obs}

\begin{figure*}[ht]
  \centering
  \includegraphics[width=1.0\textwidth]{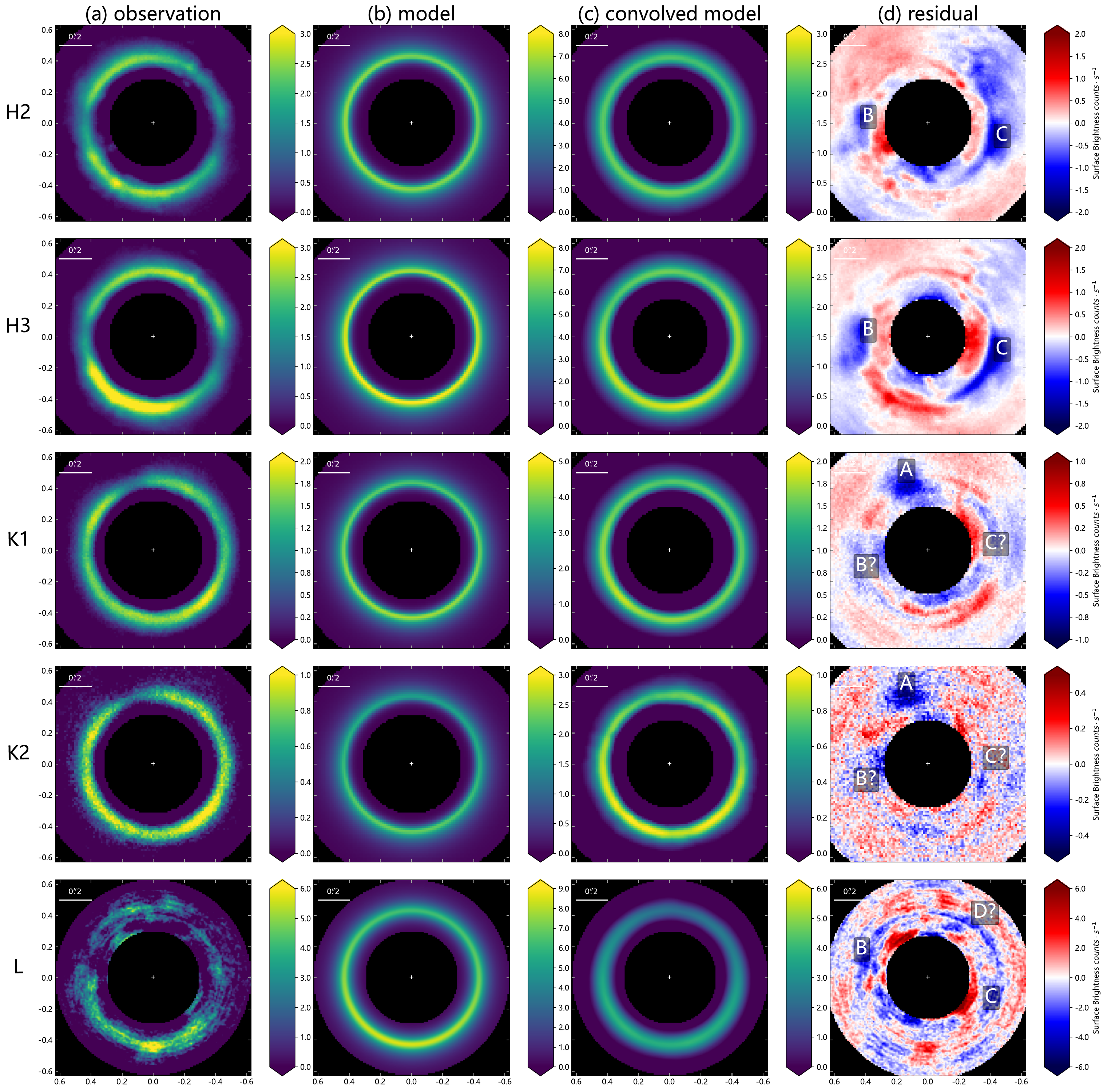}
  \caption{(a) The observation of J1604 after subtracting the stellar speckles. (b) The best-fitting model.  (c) The corresponding best-fitting model convolved with PSF and subtracted by the stellar speckles  components. (d) The residual images after removing the best-fitting model and the stellar speckles components, see Sect.~\ref{sec-forward}. The corresponding bands from top to bottom images are $H2$ and $H3$ (2015 Aug 13), $K1$ and $K2$ (2015 Jun 10), and $L$ (2017 May 10). Numerical labels indicate  the position of potential shadow features in the disk in Sect.~\ref{sec-shadow}.} \label{fig:fig-all}
\end{figure*}

\subsection{Keck/NIRC2} % proposal data

We conducted $L'$-band observation of J1604 using the Keck/NIRC2 vortex coronagraph, which has a pixel size of 9.942 mas (e.g., \citealp{Service2016}; \citealp{Mawet2019}), on UT 2017 May 10  (Proposal ID: C280; PI: D.~Mawet). The total integration time was 3285 s (=1 s $\times$ 45 coadds $\times$ 73 frames). During the observation, the  parallactic angle rotation was $36\fdg2$.
In this work, we derotated the images to align the disk and then median-combined the different exposure frames, and thus field rotation (i.e., parallactic angle change) allows for quasi-static noise removal. %So, the parallactic angles rotation could avoid the fake signal of the detector.
The central wavelength of the  $L'$-band filter is 3.776$~\mu$m. 
The observing list consists of seven targets including J1604 and six other stars which we used as reference stars to remove the star light and speckles for J1604. 
%including 2MASS~J19121875-2137074, specifically selected for its similar brightness and elevation as J1604. 
%a specifically selected reference star, 2MASS~J19121875-2137074, for its similar brightness and elevation as J1604. 
%Five additional targets were observed as reference stars.
%Additionally, corresponding point-spread function (PSF) was obtained with the star positioned outside the coronagraphic mask using a neutral density filter ND2, whose  exposure time is 0.75s (0.0075 s $\times$ 100 coadds $\times$ 1 frames).
The corresponding off-axis point spread function (PSF) was obtained by positioning the star outside the coronagraphic mask with a %neutral density filter ND2 and a 
total exposure time of 0.75s. %(0.0075 s $\times$ 100 coadds $\times$ 1 frames).

We pre-processed the observation data by performing flat-fileding, bad pixel correction, and background removal \citep{Xuan2018AJ}. %\sout{The reference images were selected based on their cosine similarity against the pre-processed target images.}
%values between these reference images and the images of J1604. 
For post-processing, we adopted reference differential imaging (RDI): we used the Karhunen--Lo\`{e}ve image projection algorithm \citep[KLIP:][]{Soummer2012} with 6 components to maximize speckle removal while presenting surface brightness and structure. %The reference images were selected based on their Cosine similarity values against the pre-processed target images. 
%\textcolor{red}{
The reference images were chosen from the exposure of the reference stars, selected on the basis of their cosine similarity against the pre-processed target images.%}
To generate the 6 components, we used 73 reference images, which match the number of frames of the J1604 observation. To obtain the final result, we de-rotated  all the images to north-up and east-left based on their parallactic angles before median-combining  them altogether, see Fig.~\ref{fig:fig-all}(a).

\subsection{VLT/SPHERE} % archival data

We retrieved archival $H$-band infrared dual-band imager and spectrograph \citep[IRDIS;][]{IRDIS_Dohlen2008SPIE} observations of J1604 on 2015 August 13 under program (ID: 295.C-5034) and the $K$-band IRDIS observations data on 2015 June 10 under program (ID: 095.C-0673). IRDIS is a dual-band imager (DBI; \cite{Vigan2010}) in SPHERE that can produce simultaneous images at two nearby wavelengths, such as $H2$: 1.593$~\mu$m and $H3$: 1.667$~\mu$m, or $K1$: 2.110$~\mu$m and $K2$: 2.251$~\mu$m\footnote{\url{https://www.eso.org/sci/facilities/paranal/instruments/sphere/inst/filters.html}}. The pixel size of IRDIS is 12.25 mas \citep{Maire2016}. During the observation, the parallactic angle rotation in $H$-band was  $91\fdg9$, and $140\fdg2$ in $K$-band. Each frame's exposure time of $H$-band and $K$-band is 32s. Additionally, corresponding non-coronagraphic stellar PSF was obtained with a neutral density filter ND2 and an exposure time of 0.8375s. We employed the reference library from \citet{Xie2022} to improve the performance of RDI. 

We pre-processed the raw data using the \texttt{vlt-sphere}\footnote{\url{https://github.com/avigan/SPHERE}, version 1.4.2.} pipeline \citep{Vigan2020ascl} to apply dark, flat, and bad pixel corrections to both the coronagraphic images and the non-coronagraphic stellar PSF image. The images were aligned using the satellite spots. Similarly to the procedure employed for the NIRC2 $L'$-band observations, we selected the reference images exhibiting the highest correlation with the images of J1604, and image numbers match the number of frames of the target observation data. We then performed data reduction using  KLIP. We de-rotated the images based on their parallactic angles and then median-combined the all frames in each filter. The resulting images are shown in Fig.~\ref{fig:fig-all}.

%Moving shadows: \citet{Debes2023} see also \citet{MuroArena2020}

\section{Analysis and Results}\label{sec-ana}
\subsection{Forward Modeling}\label{sec-forward}
To recover the surface brightness of J1604 disk in the 5 filters, we adopted the forward modeling technique \citep[e.g.,][]{mazoyer20}. In one band, we injected a negative disk model to the pre-processed data, then performed KLIP with reference differential imaging to minimize the residuals. The reduction parameters are identical to Sect.~\ref{sec-obs}.

%We employed  the static geometric disk models. %from \citet{Millar2015ApJ} 
We used a static and geometric disk model to analytically describe the spatial distribution of the scatterers within the disk: we adopted the three-dimensional distribution function in cylindrical coordinates from \citet{Augereau1999}:

%from Following the approach of \citet{Augereau1999},  with a three-dimensional function in cylindrical coordinates:
%following the approach outlined by ,
\begin{equation}\label{eq-disk}
n (r,z) \propto{\left[\left(\frac{r}{r_{\rm c}}\right)^{-2\alpha_{\rm in}}+ \left(\frac{r}{r_{\rm c}}\right)^{-2\alpha_{\rm out}}\right]}^{-\frac{1}{2}}\exp\left[-\left(\frac{z}{hr}\right)^{2}\right] ,
\end{equation}
where $h$ is the scale height, $r_{\rm c}$ is the critical radius, $\alpha_{\rm in}\textgreater0$ and  $\alpha_{\rm out}<0$  are the asymptotic power law indices when $r \ll r_{\rm c}$ and $r \gg r_{\rm c}$, respectively. We adopted $h=0.04$ from the vertical structure study of \citet{bault2009}, since it is not constrained  for face-on systems and we only focus on the surface brightness of the disk in this study.

Two specific angle values of the disk are needed in the modeling process, the inclination angle and the position angle.
The inclination angle ($\theta_{\rm inc}$) is defined as the dihedral angle between the disk mid-plane and the sky and the position angle ($\theta_{\rm PA}$) is defined as the position angle of the disk's semi-major axis measured from North to East.
%We use the position angle of the disk's semi-major axis to represent the position angle of the disk ($\theta_{\rm PA}$), which is measured from North to East. 
This semi-major axis is chosen as the one $90^{\circ}$ counterclockwise from the semi-minor axis that is closer to Earth.
%defined as the 
%position angle of the disk's semi-major axis---the one which is $90^{\circ}$ counterclockwise from the semi-minor axis that is closer to Earth---measured from North to East. 
% the position angle of the disk ($\theta_{\rm PA}$) is defined as the position angle of the disk's semi-major axis---the one which is $90^{\circ}$ counterclockwise from the semi-minor axis that is closer to Earth---measured from North to East. 
For these two angles, we adopted fixed values from the analysis of ALMA observation data by \citet{Dong2017} and \citet{Stadler2023}, with $\theta_{\rm inc} = 6^\circ$ and $\theta_{\rm PA} = 258\fdg7$.
%In consideration of the low signal-to-noise ratio (S/N), we adopted fixed values for these two angles from \citet{Stadler2023, Dong2017}.  

The scattering angle is defined as the angle measured from the incident light ray to the outgoing ray. The intensity of scattered light as a function of scattering angle is referred  as scattering phase function (SPF). In this study, we adopt the parametric SPF in \citet{hg41} in total intensity:

\begin{equation}\label{eq-hg}
I_{\rm tot} (\theta) = \frac{1-g^2}{4\pi (1+g^2-2g\cos\theta)^{3/2}},
\end{equation}
where $\theta$ is the scattering angle, and  $g\in (-1, 1)$ is the scattering asymmetry parameter with $-1 < g < 0$ for backward scattering and $0 < g < 1$ for forward scattering.
%where $\theta$ is the scattering angle and $g \in [-1,1] $  is a parameter that ranges from backward scattering when $g \leq 0$ to forward scattering when $g \geq 0$.

We combined the static geometric disk model with their corresponding scattering phase function to obtain a model image. To match the brightness of the disk model image with the observation data, we introduced a multiplicative scaling factor,
which was raised to powers of 10 then multiplied with the disk model images.%\textcolor{red}{denoted as $f_{\rm flux }$ }. 
In this work, we used the \texttt{DebrisDiskFM} package \citep{ren19} for disk image modeling using the \citet{Millar2015ApJ} codes which have been adopted in modeling ring-like structures in debris disks and protoplanetary disks \citep[e.g.,][]{Wang2020AJ, Quiroz2022ApJ, Ren23debris}. Recognizing the potential misalignment between the image center of simulated data and observation data due to the instrument jitter or inaccuracies in the pre-processing procedures, we adopted two parameters $\Delta x$ and $\Delta y$ to represent the shifts of the disk along the south-north direction and east-west direction, respectively. Due to the low  S/N of the $L'$-band data, we set both shifts to zero when modeling the $L'$-band data. 

%To obtain a better fitting result, we decided to use the forward modeling technique, where a negative disk model was added to the pre-processed data and then KLIP reduction to minimize the residuals. 
To produce an observed disk model through finite telescope aperture, we rotated the model image generated from \texttt{DebrisDiskFM} package based on the parallactic angle corresponding to each of the observation images and convolved it with the normalized PSFs obtained in Sect.~\ref{sec-obs}. 
%The rotated model image was convolved with the PSF obtained in Sectin 2.
%The next step involved convolving the model image with PSF. 
Given the distinctive characteristics of the NIRC2 instrument, we also applied a transmission map of the NIRC2 vortex coronagraph \citep[e.g.,][]{Wang2020AJ, Quiroz2022ApJ} during the $L'$-band modeling. 
We then subtracted the disk model images from the target images (i.e., negative injection) based on the corresponding parallactic angles, and removed the stellar speckles using KLIP (see Sect.~\ref{sec-obs}) to obtain the residual images. 
%The reference images were selected based on their Cosine similarity values against the pre-processed target images. 
%During the generation of 6 components with KLIP method, the number of selected reference images is chosen to match the number of frames of the target observation data.  
%A backward rotation was applied to the residual images in accordance with their parallactic angles before 
We calculated the element-wise median and standard deviation of the derotated residuals to obtain the final result for negative injection.% and the standard deviation value of each pixel of the residual images. 
%Next, we calculated both the median value and the standard deviation value of each pixel using the residual images. 

The residual images have significantly smaller standard deviation values in the edge region, we thus excluded the edge region in our analysis.
%because of their significantly smaller standard deviation values.
%Notably, during this process, we omitted the edge region where the standard deviation was significantly small. 
%Finally, we computed the loglikelihood function \ref{eq-lnlike} assuming independent Gaussian distribution for the SPHERE pixels: 
To obtain the best-fit models for the observed data, we maximized the log-likelihood function assuming independent Gaussian distribution for the final negative injection result:
\begin{equation}\label{eq-lnlike}
\ln  L (\Theta \mid X_{\rm obs})= -\frac{1}{2} \sum_{i=1}^N \left(\frac{X_{{\rm res},i}}{\sigma_{{\rm res},i}}\right)^2
-\sum_{i=1}^N\ln\sigma_{{\rm res}, i}- \frac{N}{2}\ln(2\pi) ,
\end{equation}
where $\Theta$ is the set of the disk and offset parameters (i.e., $\theta_{\rm inc}, \theta_{\rm PA}, \alpha_{\rm in}, \alpha_{\rm out}, r_{\rm c},g, f_{\rm flux}, \Delta x,\Delta y$), $X_{{\rm res}}$ is the element-wise median derived from residual images across diverse exposure frames%of the residual images from different exposure frames
with $N$ pixels, $\sigma$ is the uncertainty map that has the same dimension as $X$.
To obtain the best-fitting disk parameters for the observational data and explore the parameter space, we employed the \texttt{emcee} package \citep{Foreman2013}.
In order to reduce the influence of shadows of the disk \citep{Pinilla2018} during the modeling of the disk, we selectively excluded specific regions when calculating the likelihood function. This exclusion was achieved through the application of a distinctive mask as depicted in Fig.~\ref{fig:mask-all}.

Considering the similarity between the $H2$ and $H3$ band, and that  signal-to-noise (S/N) of $H3$ band is higher than $H2$ band, we fixed the parameters $\Delta x~({\rm pixel}), \Delta y~({\rm pixel}) $ of $H2$-band  to the best parameters derived from the $H3$ band. Similarly, in the $K$ bands, where S/N of $K1$ band better than $K2$-band, we fixed the parameters $\Delta x~({\rm pixel}),\Delta y~({\rm pixel})$ of $K2$-band  to  the optimal  parameters obtained from $K1$-band.  We presented the 50$\pm$34th percentiles for the retrieved disk parameters in Table~\ref{tab:model-para}. We present and discuss the MCMC modeling results in appendix~\ref{sec-corner}.

\begin{table*}
\setlength{\tabcolsep}{14pt}
\renewcommand{\arraystretch}{1.3}
\caption{Best-fitting parameters for J1604 disk in total intensity in scattered light\label{tab:model-para}}
\begin{tabular}{c c r r r r r} \hline\hline
 parameter  & unit & $H2$ band & $H3$ band & $K1$ band  & $K2$ band & $L'$ band\\ \hline
$\theta_{\rm inc}\tablenotemark{a}$   & $^\circ$ & 6    & 6     & 6       & 6        & 6\\ 

$\theta_{\rm PA}\tablenotemark{b}$   & $^\circ$ & 258.7 & 258.7 & 258.7 & 258.7  & 258.7\\ 

$\alpha_{\rm in}$    & & $18.578^{+2.027}_{-1.684}$ & $25.064^{+1.581}_{-1.450}$  & $21.968^{+2.573}_{-1.490}$   &$21.063^{+2.714}_{-2.486}$ & $11.493^{+2.982}_{-2.256}$\\ 

$\alpha_{\rm out}$    & & $-6.545^{+0.271}_{-0.271}$ & $-6.161^{+0.090}_{-0.101}$  & $-7.322^{+0.212}_{-0.204}$ &$-6.856^{+0.402}_{-0.427}$ & $-8.986^{+1.405}_{-1.694}$\\

$r_{\rm c}$    & au & $60.507^{+0.341}_{-0.340}$ & $59.993^{+0.154}_{-0.147}$  & $61.682^{+0.238}_{-0.260}$   & $61.872^{+0.430}_{-0.385}$ & $65.727 ^{+1.426 }_{-1.240}$\\

$g$    & & $0.126^{+0.041}_{-0.046}$ & $0.277^{+0.026}_{-0.024}$ & $0.307^{+0.021}_{-0.020}$   & $0.717^{+0.162}_{-0.183}$ & $0.743^{+0.173}_{-0.200}$\\

$\Delta x$\tablenotemark{c}     & pixel & $0.053$ & $0.053^{+0.031}_{-0.031}$ & $1.12^{+0.05}_{-0.04}$   & 1.12& 0\\

$\Delta y$\tablenotemark{c}     & pixel & -1.279 & $-1.279^{+0.037}_{-0.038}$ & $-0.26^{+0.06}_{-0.07}$   & -0.26& 0\\ \hline

\end{tabular}
\begin{flushleft}
\tiny  \textbf{Notes}:

$^a$ {$\theta_{\rm inc}$ was fixed inferred from the dust continuum \citep{Dong2017}.}

$^b$ {$\theta_{\rm PA}$ was fixed inferred from  the $^{12}$CO  intensity channel maps of the disk \citep{Stadler2023}. }

$^c$ {Offset between ring center and star. $\Delta x$ and $\Delta y$ \citep{Millar2015ApJ} expressed in units of instrument pixels, are fixed for specific bands. For $H2$-band, we  employed the best-fitting model from the $H3$ band; for $K2$-band, we utilized the best-fitting from the $K3$-band; and for the $L'$-band, no offset adjustments were applied.}
\end{flushleft}
\end{table*}

\begin{figure}[htb]
  \centering
  \includegraphics[width=0.5\textwidth]{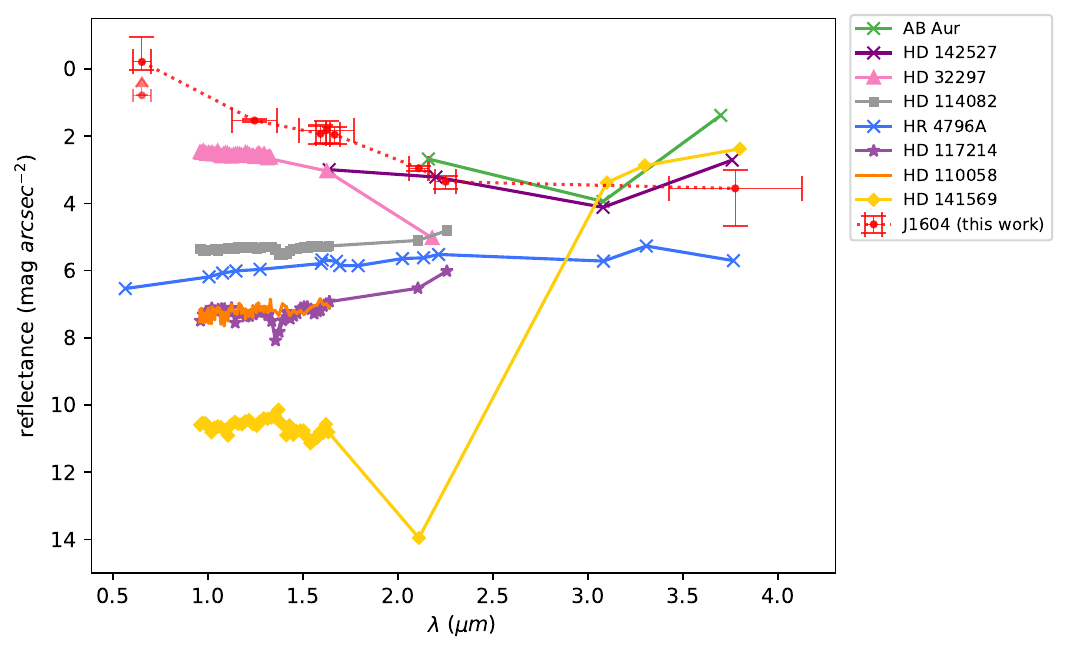}
  \caption{The reflectance  of different bands and different disk including HR~4796A \citep{Debes2008, Rodigas2015, Milli_2017}, HD~32297 \citep{Bhowmik2019}, AB~Aur \citep{Betti2022AJ}, HD~142527 \citep{Honda2009ApJ}, HD 141569 \citep{Singh2021A&A, Kueny2024}, HD~110058 \citep{Stasevic2023A&A}, HD~114082 and HD~117214 \citep{Engler2023A&A}. We scaled these disk with  the square of their radial distances ratio to J1604, see Sect.~\ref{sec-comparison}. 
  %In the Ks bands, the black and blue points are overlapped with each other. 
  %We presented the relative reflectance values expressed in units of mag ${\rm arcsec}^{-2}$, which is calculated using $-2.5 \log_{10} \rho$ and $\rho$ is the relative reflectance in Eq.\eqref{eq-err}. 
  For J1604 in $R$-band, we only provided a lower limit for the total intensity reflectance using polarization data. Note: due to stellar activities, and thus the lag in photon arrival times between the star and the disk, the accuracy of the reflectance measurement is limited, see Sect.~\ref{sec-Stellarvary}.
  } \label{fig:reflectance}
\end{figure}

\begin{figure*}[htb]
 \centering
 \includegraphics[width=0.95\textwidth]{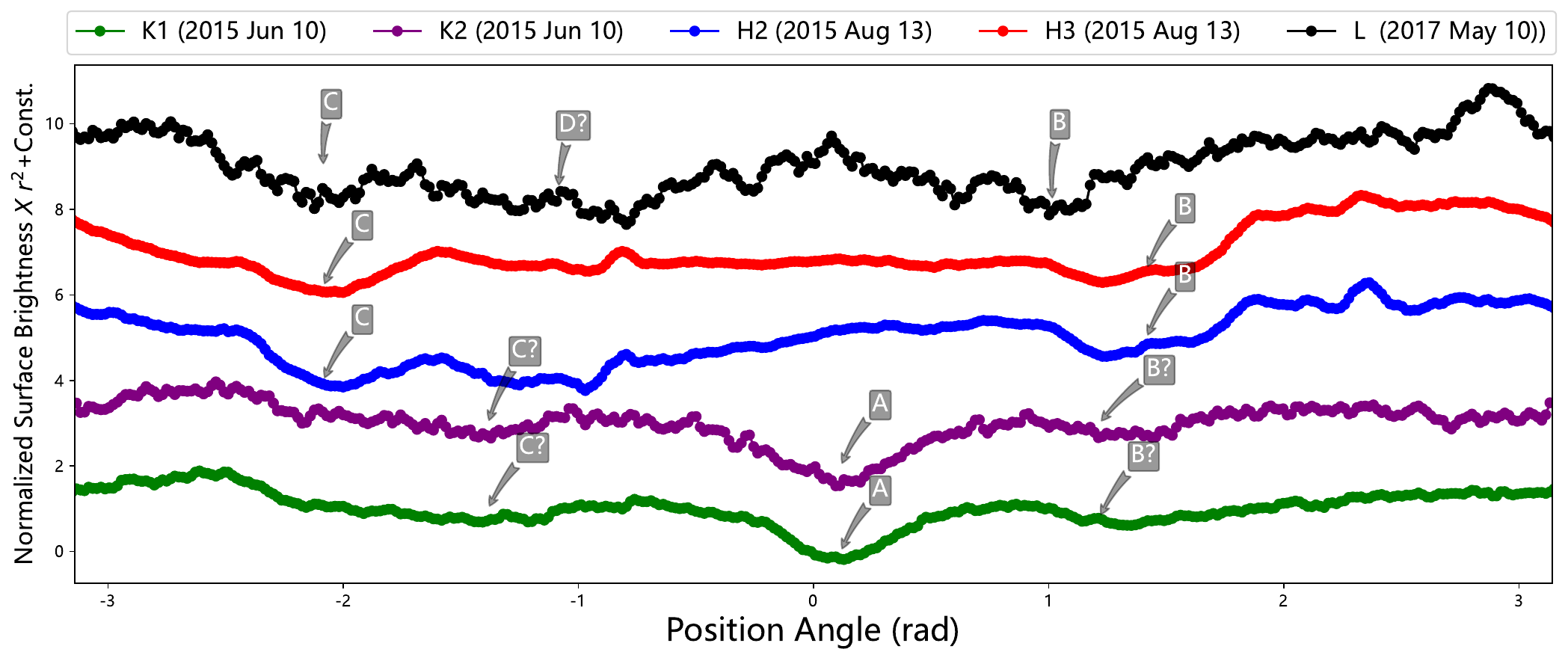}
 \caption{
 %After employing the KLIP method to remove stellar contributions and , the azimuthal profile of deprojected J1604 observation data %subsequent to the removal of stellar contribution using the KLIP method %Azimuthal profile of deprojected J1604 observation data after the removal of stellar contribution with KLIP method (\textbf{after the removal of stellar contribution: to be specified by Huisheng}) 
 %was scaled by $r^2$ illumination correction. 
 To derive the azimuthal profile, we deprojected J1604 observation data, which was processed to remove stellar contributions using the KLIP method and then scaled by $r^2$ illumination correction. Component `A' vanished between 2015 Jun and Aug, while `B' and `C' enhanced then. One additional component `D' might start appearing in 2017 May. %For ease of presentation, we normalized the profiles and assigning distinct offsets to various bands observation in the y-axis direction, positioned the `$A$' shadow at the center. 
 The vertical arrangement, from bottom to top, aligns with the chronological sequence the observation period.}     
 \label{fig:fig-brightness_R2scale}    
\end{figure*}

\subsection{Aperture photometry}\label{sec-Aperture}

The effective telescope aperture of Keck II is equivalent to that of a circular aperture with a diameter of 9.96 meters.\footnote{\url{https://www2.keck.hawaii.edu/observing/kecktelgde/ktelinstupdate.pdf}} %\citep{Keck2002}
The aperture of SPHERE is 8.0 meters in diameter seen from SPHERE \citep{rupprecht2005requirements}. We obtained the host star flux, denoted as $F_{\rm star}$ (units: $\rm counts~{\rm s}^{-1} $), by integrating the image counts within the first-order diffraction ring divided by the exposure time. 
%To approximate an integer integral radius, we selected a ratio of 1.258 as the aperture size to $\frac{\lambda}{D}$ in different bands. 
We set the aperture size to 1.22$\frac{\lambda}{D}$ across all bands, to ensure that the ratio of the calculated flux to the true flux remains the same across different bands. 
After model fitting, we obtained a Markov chain of corresponding parameters and used it to create disk model images. 
%The average disk flux, denoted as $F_{\rm disk}$, was derived by median-combine the surface brightness within the same aperture used for stellar flux measurement. 
%The average disk flux, denoted as $F_{\rm disk}$, was derived by median-combine the model image counts per pixel surface brightness within the same aperture used for stellar flux measurement.
%The average disk flux, denoted as $F_{\rm disk}$, was derived using the median pixel count number of the model image within the same aperture, divided by the angular size of a pixel and the exposure time.
We obtained average disk flux $F_{\rm disk}$ (units: $\rm counts~{\rm s}^{-1} ~ {\rm arcsec}^{-2}$) using the median pixel count number of the model image within the same aperture divided by the angular size of a pixel and the exposure time.
The center of aperture was placed near the position of peak surface density along the disk major axis, determined by the best-fitting disk parameter $r_{\rm c}$ and $\theta_{\rm PA}$, $\alpha_{\rm in}$, $\alpha_{\rm out}$ using Eq.~\ref{eq-disk}. 
%The aperture was placed at the position of the nearly peak surface density along the major axis of the disk determined by the best-fitting disk parameter $r_{\rm c}~(\rm au)$ and $\theta_{\rm PA}$. 
We chose this region, where the scattering angle is nearly $90^{\circ}$, to mitigate the influence of the phase function and facilitate comparisons with studies for other systems. 
%The unit of $F_{\rm disk}$ was converted into $\rm counts~{\rm s}^{-1} ~ {\rm arcsec}^{-2}$ by dividing the angular size of a pixel and the exposure time.  
%We converted  the unit of $F_{\rm disk}$ into $\rm counts~{\rm s}^{-1} ~ {\rm arcsec}^{-2}$ by dividing the angular size of a pixel and the exposure time.  
The resulting ratio of $F_{\rm star}$ to $F_{\rm disk}$ corresponds to the reflectance $\rho$, expressed in units of $  \; {\rm arcsec}^{-2}$. This transformation effectively eliminates the influence of  the stellar color and the instrument effects, enabling the quantification of the disk's reflectance $\rho$ as a function of wavelength.  %Considering the gain, we convert edu units to photon count.

Similarly we conducted  photometry on the observation data of the disk. However, it's important to note that the convolution effect of the instrument and the KLIP  algorithm would influence  the reflectance by introducing  image blurring and reducing the surface brightness of the disk. 
To mitigate these effects, we divided the KLIP-reduced image by the best-fitting disk model image and subsequently mean-combined the resulting image within the region of the disk, excluding the shadows (see Fig.~\ref{fig:fig-all}), to determine the  throughput.  
We then corrected the reduced image by dividing it by the throughput and  performed photometry on the corrected 
image, in this way, we can reduce the possibility that the disk model in Eq.~\ref{eq-disk} is not an accurate description of J1604 disk morphology in calculating the reflectance. This additional photometry step also  allowed us to assess  the accuracy of the reflectance obtained from the disk model (see  Fig.~\ref{fig:J1604reflectance}). When the values of the  reflectance respectively obtained from the corrected image and the model image are in close agreement, we consider the results reliable.

%The gain of H-band observation is $1.75e^{-}/ADU $and for K-band is $1.8e^{-}/ADU $, for L-band is $4e^{-}/ADU$.
The band widths of $H2$-band, $H3$-band, $K1$-band, $K2$-band and $L'$-band are 0.052, 0.054, 0.102, 0.109~$~\mu{\rm m}$ \citep{SPHEREManual}, and 0.7$~\mu{\rm m}$\footnote{\url{https://www2.keck.hawaii.edu/inst/nirc2/filters.html}}, respectively. 
Furthermore, we calculated the errors in reflectance $\rho$ by combining the uncertainties  of the stellar flux  and the disk flux:

%as detailed in Equation \ref{eq-err}.
\begin{equation}\label{eq-err}
\begin{cases}
\displaystyle  \quad \rho &= \displaystyle \frac{F_{\rm disk}}{F_{\rm star}} \\[10pt]
\displaystyle \left(\frac{\delta \rho}{\rho}\right)^2&= \displaystyle \left(\frac{\delta F_{\rm disk}}{F_{\rm disk}}\right)^2 + \left(\frac{\delta F_{\rm star}}{F_{\rm star}}\right)^2
\end{cases}
\end{equation}
where $\delta F_{\rm disk}$ for modeled disk image was calculated using the Markov chain of model fitting. %with $1 \sigma$ significance 
We corrected  the residual image, as shown in Fig.~\ref{fig:fig-all}, by dividing it by the throughput and  performed photometry on the corrected  image out of the disk region to obtain $\delta F_{\rm disk}$.

%\sout{Doing} 
We provided detailed aperture photometry results in Table~\ref{tab:result-aperture}. The final results, as shown in Fig.~\ref{fig:J1604reflectance} and Fig.~\ref{fig:reflectance}, reveal  that  the disk scatters less light in longer wavelengths compared to shorter wavelengths. By analyzing the reflectance at various  wavelengths, future work might constrain physical properties of the dust population, including size, shape and composition of the micron-sized dust particles. %Moreover, conducting an analysis of the disk's composition has the potential to aid in the determination of  its bulk compositions.

\begin{table*}
\setlength{\tabcolsep}{1pt}
\renewcommand{\arraystretch}{1.}
\caption{Detailed values for disk surface brightness measurement\label{tab:result-aperture}}
\begin{tabular}{c c r r r r r r r r} \hline\hline
 &unit  & $R$ band & $J$ band & $H$ band& $H2$ band & $H3$ band & $K1$ band  & $K2$ band & $L'$ band\\ \hline
Epoch & & 2015.06 & 2017.08 & 2016.06 & 2015.08 & 2015.08 & 2015.06  & 2015.06 & 2017.05\\  
Wavelength &$~\mu{\rm m}$& $0.65\pm 0.05$& $1.24\pm 0.12$& $1.62\pm 0.14$& $1.59\pm 0.02$& $1.67\pm 0.02$& $2.11 \pm 0.05$ &$2.25 \pm 0.06$& $3.78 \pm 0.35 $\\
Resolution\tablenotemark{a}   &$\rm mas$& 20 & 39 & 51 & 50& 52& 66 &71 & 95\\ 
$F_{\rm star}  \tablenotemark{b}$    & $10^3~\rm counts~{\rm s}^{-1}$ &$27\pm 1$ &$367\pm 8$ &$524\pm 77$ &$190\pm 42$ & $211\pm 45 $   & $263\pm 4 $  & $165\pm 2 $ & $1110\pm 1 $\\

$F_{\rm disk}^{\rm (model)}$\tablenotemark{c}&$10^2~\rm counts~{\rm s}^{-1} ~ {\rm arcsec}^{-2}$& $131 \pm9$\tablenotemark{f}& $ 892 \pm 28$ & $ 980\pm 250$ & $ 322^{+6}_{-5}$ & $ 348^{+3}_{-3}$ & $ 173^{+3}_{-2}$   &$ 75^{+2}_{-2}$ & $ 420^{+31}_{-31}$ \\%median-combined

$F_{\rm disk,\ 1}^{\rm (obs)}$\tablenotemark{d} &$10^2~\rm counts~{\rm s}^{-1} ~ {\rm arcsec}^{-2}$& & &&$ 160 \pm 40 $&$170 \pm 26 $&$ 170 \pm 12 $&$ 93 \pm 13 $ &$190 \pm 270$\\%median-combined

$F_{\rm disk,\ 2}^{\rm (obs)}$\tablenotemark{e} &$10^2~\rm counts~{\rm s}^{-1} ~ {\rm arcsec}^{-2}$&&&&$ 384 \pm 35 $&$358 \pm 26 $&&&$470 \pm 270$\\%median-combined
\hline

\end{tabular}
\begin{flushleft}
\tiny  \textbf{Notes}:

$^a$ {Spatial resolution obtained from $1.22\lambda/D$, where $\lambda$ is the central wavelength and D is the telescope pupil size seen by the instrument.}
%$^b${Considering the different pixel size of different instrument the $\frac{\lambda}{\rm D}(\rm mac)$ convert to corresponding pixel unit.} 
%$^c${For convenience, we round it up.}

$^b$ {In both the $H$-band and 
$K$-band, two stellar flux images were acquired on separate nights. The mean flux in the $K$-band was computed based on the two observed values, with preference given to the flux from the second night, accounting for seeing conditions. And we calculated the standard deviation with two flux values the $H$-band and $K$-band considering potential stellar brightness variation. And we also computed  $\sqrt{n}$ as
the photonic noise following Poisson distribution. For instance, in the case of observations in the $H2$ band, the photonic noise was computed as $\sqrt{263\times1000\times0.837}/0.837$. The overall errors in stellar flux were determined by combining the photonic noise and the standard deviation with two night flux values. }

$^c$ {Average disk surface brightness for the unconvolved models. And the error was calculated  by the Markov chain of corresponding disk parameters obtained in the process of forward modeling.}

$^d$ {Average disk surface brightness for observation with throughput correction in the major axis. And the error was calculated by the residual image in the region out of the disk.}

$^e$ {Average disk surface brightness for observation with throughput correction out of major axis. Considering the low inclination of the disk, the scatter angle of the region  is nearly $90^{\circ}$.}

$^f$ {The average polarized disk surface brightness for the unconvolved disk model.}
\end{flushleft}
\end{table*}

\section{Discussion}\label{sec-dis}

\subsection{Stellar brightness variation}\label{sec-Stellarvary}
%The host star of J1604 is a young and active star exhibiting frequent flare activities. Additionally, the host star of J1604, often referred to as a  young 'dipper' star, exhibits quasi-periodic or aperiodic dimming events in its optical and infrared light curves. 
J1604 is a young active `dipper' star, which exhibits quasi-periodic or aperiodic dimming events in its optical and infrared light curves. 
An analysis of the three known dippers with publicly available resolved sub-mm data, conducted by \citet{Ansdell2016MNRAS}, concluded that nearly edge-on viewing geometries of the outer disk could not explain the depth of light curves. 
%boma, 光变是内辐射源光变。
\citet{Sicilia2020} studied 10 deep eclipses of J1604 and found the brightness variation in $H$-band reached 0.2 to 0.7 mag.
%magnitudes ranging from 0.2 to 0.7 and detected at least 10 deep eclipses during their observations, all of which have been confirmed in multiple bands. 
In the $H2$- and $H3$-band observations here, the stellar flux density on the second night is nearly twice that of the first night. The seeing conditions on the first night was $1\farcs65 \pm 0\farcs01$, while on the second night, they were $1\farcs29 \pm 0\farcs05$. This discrepancy indicates the stellar flux observed on the first night was influenced by adverse weather conditions and may not be accurate. Therefore we adopted the stellar flux obtained on the second night, and conservatively used the standard deviation of the two measured flux values as the error associated with the stellar flux measuremeant. In other bands, we did not detect similar stellar flux variations.
%Such a large variation of the stellar flux meant one possible dimming event happened during our observation, and could affect the match of the reference star resulting bad performance of the classical PSF subtraction \citep{Stark2023ApJ}. However, in our measurements, the presence of a diverse reference star library helped mitigate the influence of the temporal variations in the stellar flux on the KLIP result. %obtained through the KLIP method. 

The actual stellar flux variation could impact the measurements of reflectance because the photons emitted from the star arrives at the surrounding disk with a delay of several hours to a few days. For J1604, this delay is about 8 hours: the scattered light flux variation of the disk lags behind the flux variation of the star. Any stellar flux variation happened during the observation could lead to measured fluctuations in the disk's brightness that deviate from actual values, biasing the derived reflectance measurements. 
In the future, with the aim of enhancing the reliability of the reflectance of J1604, more careful treatment of the difference in light arrival-time should be performed \citep[e.g., TESS:][]{Stark2023ApJ}.

%\citet{Stark2023ApJ} observed temporal variations in the  flux of the host star, which consequently   changed the $(B - V)$ color index of HD 53143. 
%These variations  notably  affected the match of the reference star, subsequently reducing  the performance of the classical PSF subtraction. 
%However, in our measurements, the presence of a diverse reference star library helped mitigate the influence of the temporal variations in the star's flux on the result obtained through the KLIP method. But it's essential to acknowledge that when the host star experiences brightness changes, the emitted light hits the surrounding disk after a delay of several hours. As for the J1604, the delay timescale was about 8 hours. Consequently  the scattered flux variation of the disk lags behind  the flux variation of the star. 

%Additionally, we acquired stellar flux measurements on the initiation and/or end of the observation, and median-combined the frames of the disk obtained from longtime observation, the duration of which was shorter than the delay timescale, to obtain . 

%Any potential  dimming event happened during the observation could lead to measured fluctuations in the disk's brightness that deviate from actual values, affecting the derived reflectance measurements. So in the future, with the aim of enhancing the reliability of our conclusions, we plan to conduct additional observations of J1604.

\subsection{Comparison with previous works}\label{sec-comparison}

%The KLIP-method would reduce the throughput of disk, using the best-fitting model, obtained by forward-modeling, to 

%As for the H-band, because the measure areas is coincident with  the shadow, so the throughput of model is less than the observation with throughput correction.

There are a few studies focusing on measuring the broadband reflectance or reflectance spectrum for circumstellar disks. By situating the J1604 measurements in the reflectance of HR~4796A \citep{Debes2008, Rodigas2015, Milli_2017}, HD~32297 \citep{Bhowmik2019}, AB~Aur \citep{Betti2022AJ}, HD~142527 \citep{Honda2009ApJ}, HD~141569 \citep{Singh2021A&A, Kueny2024}, HD~110058 \citep{Stasevic2023A&A}, HD~114082 and HD~117214 \citep{Engler2023A&A}, we could study both the difference and similarity between the environment in the debris disk -- where giant planet formation has reached completion -- and the environment within the protoplanetary disk (e.g., AB~Aur, J1604, HD~142527, HD 141569) where giant planet formation is currently in progress. 
%The reflectance spectrum of HR~4796A, as investigated by \citet{Milli_2017}, encompassed a broader wavelength range, extending from 0.5 to 4 microns and showed a monotonic increase trend from 0.5 to 1.5 microns. However, it kept nearly constant from 1.5 to 4 microns. 

%不太懂这个颜色。
%Additionally, the reflectance spectrum of HD~32297, as explored by \citet{Bhowmik2019}, was limited to a narrow wavelength range, spanning only from 1 to 1.8 microns. 
%This spectrum exhibited  a monotonically decreasing trend with wavelength, giving the disk a gray to blue color in the YJH spectral range.   

Existing studies on circumstellar disks showed %\sout{a few cases without clear trends}
several features on their reflectance. \citet{Milli_2017} measured the reflectance of HR~4796A from 0.5$~\mu$m to 4.0$~\mu$m, which showed a monotonic increase trend from 0.5$~\mu$m to 1.5$~\mu{\rm m}$ and then almost kept constant from 1.5$~\mu$m to 4$~\mu$m. 
\citet{Bhowmik2019} measured the reflectance spectrum of HD~32297 from 1 to 1.8~$~\mu$m, the spectrum exhibited a monotonically decreasing trend with wavelength, which renders the disk a gray to blue color in the $YJH$ spectral range. 
%\sout{Employing a color-color difference diagram citep{Inoue2008PASJ}, citet{Betti2022AJ} discerned the existence of icy grains on the surface of the disk of AB~Aur with dust size of ${\sim}3~\mu{\rm m}$.} 
Employing a color-color difference diagram \citep{Inoue2008PASJ}, \citet{Honda2009ApJ} discerned the existence of silicate and H2O ice grains of  on the surface of the disk of HD~142527 with dust size of ${\sim}1~\mu{\rm m}$. Simiraly, AB Aur and HD~141569 was observed a absorption at the ice line at ${\sim}3.1~\mu{\rm m}$ \citep{Betti2022AJ, Kueny2024}.
\citet{Engler2023A&A} found the measured reflectance spectrum of HD~117214 disk maybe indicates the presence of CO$_2$ ice as a constituent of debris material at the investigated radial distance from the star and a red color at longer wavelengths than 1.66 µm for both HD~117214 and HD~114082.

The reflectance measurements of J1604 here encompass the broadest wavelength range from 0.5$~\mu$m to 4$~\mu$m to date (e.g., HR~4796A: \citealp{Debes2008}). Considering  decrease of stellar light intensity with  distance which scales as $1/r^2$, where $r$ is the stellocentric distance, we scaled these disk reflectance with the square of their radial distances ratio to J1604 (e.g., the scaling factor for HD~114082 is $r^{2}_{\rm HD114082}/r^{2}_{\rm J1604} = (31~{\rm au}/61~{\rm au})^2 = 0.26$, and HD~32297 $r^{2}_{\rm HD~32297}/r^{2}_{\rm J1604} = (134.7~{\rm au}/61~{\rm au})^2 = 4.88$). The reflectance amplitudes of the debris disk after scaling, including HR~4796A and HD~117214 and HD~114082, were at$~10^{-2}$--$10^{-3} {\rm arcsec}^{-2}$, signifying a lower magnitude compared to the reflectance amplitude of protoplanetary disks including J1604, AB~Aur, and HD~142527, which are at ${\sim}10^{-1}~{\rm arcsec}^{-2}$.

We can identify a monotonically decreasing trend in the reflectance of J1604 over the wavelength range of 0.5$~\mu$m to 2.1$~\mu$m (i.e., blue color), contrasting with the flat or gradual increase in the reflectance of HD~114082 and HD~117214, HR~4796A, HD~110058, all of which are debris disks (i.e., gray to red color).  With the distance-dependent scaling, the reflectance %\sout{amplitudes and} 
trend of HD~32297 and HD 141569 is relatively close to J1604. 
%It is worth noting that the reflectance of HD~32297 and HD 114082. 
It could be speculated that the evolution of the disk may entail a progressive transition in the reflectance spectrum shape, shifting from J1604-like characteristic blue slope to those featuring red slope resembling HR 4796A. %HD 32297, 
%and ultimately converging towards features ‘’ observed in HR 4796A.
%Lacking more data point of IR-band, 
Additionally, the reflectance of J1604 and HR~4796A both exhibit a consistent, flat trend within the wavelength range of 2.1$~\mu$m to 4$~\mu$m. In contrast, for AB~Aur, HD 141569, and HD 141527 the reflectance at ${\sim}3.8~\mu$m is larger than that at ${\sim}2.2~\mu$m %\sout{the manifests significant and distinctive variations} 
in the same wavelength span and showed potential water ice absorption. We note the necessity for additional data points of J1604 within the wavelength range of 2.1$~\mu$m to 4~$\mu$m to explore its corresponding features.
Finally, there are potential biases in compiling and analyzing the statistical results should not be disregarded, since there are inconsistencies in existing surface brightness measuring and reporting methods in previous work. Specifically, the reflectance measurement by \citet{Singh2021A&A}, along with some other studies, utilized the average flux of the whole disk, which is different from our measurement in Sect.~\ref{sec-Aperture}.
A limited quantity of available samples hinders comprehensive analysis.  
Future studies on reflectance spectra using integral field spectrograph (IFS), including SPHERE and Gemini planet Imager (GPI), will help better probe the spectroscopic trends as well as the mineralogical composition of circumstellar disks when aided with laboratory measurements \citep[e.g.,][]{Poch23}.

\subsection{Varying  disk structure}
\subsubsection{Structure}
The disk geometry structure which we obtained by forward-modeling vary with the wavelengths. Although the fitting parameters $r_{\rm c} $ and $\alpha_{\rm in}$, $\alpha_{\rm out}$ in the $H$-band closely resemble those in the $K$-band due to the proximity in wavelength, significant differences are observed in the best-fitting parameters $r_{\rm c} $ and $\alpha_{\rm in}$, $\alpha_{\rm out}$  for the $L'$-band. The significant difference in wavelength may be attributed to the spatial segregation of dust particles, which vary in size and scattering properties, potentially influenced by size-dependent forces. For example, the unitless ratio $\beta$ between radiation pressure and gravitational forces  exhibits an inverse proportionality to grain size  for grains larger than a few$~\mu{\rm m}$ \citep{Olofsson2022A&A}. Additionally, the drag, which the gas exerts on the grains, is known to be size-dependent \citep{Takeuchi2001ApJ}.

% Considering that observations at  different wavelengths probe different grain sizes,  variations in the parameter $g$ may be  attributed  to  disparities in the  physical properties of the scattering particles. More  specifically, the disk geometry which we obtained by forward-modeling  obviously vary with the wavelengths. Although  the fitting parameters $r_{\rm c} $ and $\alpha_{\rm in}$, $\alpha_{\rm out}$ in the H-band closely resemble those in the K-band due to the proximity in wavelength, significant differences are observed in the best-fitting parameters $r_{\rm c} $ and $\alpha_{\rm in}$, $\alpha_{\rm out}$  for the L-band. The significant difference in wavelength may be attributed to the spatial segregation of dust particles, which vary in size and scattering properties, potentially influenced by size-dependent forces. For example, the unitless ratio $\beta$ between radiation pressure and gravitational forces  exhibits an inverse proportionality to grain size  for grains larger than a few$~\mu{\rm m}$ \citep{Olofsson2022A&A}. Additionally, the drag, which the gas exerts on the grains, is known to be size-dependent \citep{Takeuchi2001ApJ}.  

In comparison to $L'$-band, the values of  $\alpha_{\rm in}$ in both $H$-bands and $K$-bands are notably larger, indicating a sharper inner edge.  This result suggests  the possible presence of dust particles condensed at the inner edge, such as NH$_{3}$, water ice and hydrated minerals  \citep{s19092076},\footnote{\url{https://www.nist.gov/}} which exhibit strong light-absorption characteristics in $H$-bands or $K$-bands but exhibit a less significant effect in the $L'$-band.  Another explanation for  the observed cavity with sharp edge in both the $K$-band and the $H$-band could involve the presence of a  ${\sim}15~M_{\rm Jup}$ brown dwarf orbiting at  ${\sim}15$~au \citep{Canovas2017}. It is well-established that a massive giant planet can create a cavity with sharp edge within the disk \citep{Pinilla2012A&A, Johansen2019}. \citet{Rodigas2014ApJ} derived a linear expression relating a shepherding planet's maximum mass to the debris ring's observed width in scattered light. Yet J1604 is not a debris system but a protoplanetary/transition disk which is rich in gas thus can involve more complex mechanisms. Future work on explaning the radial distribution of rings in protoplanetary/transition disks could better explain the observation in this study. %Additionally, the outer edge of J1604 disk is not sharp, indicating the different dynamics characteristic of the protoplanetary disk, where giant planet formation is currently in progress. % The wavelength-dependent variation in the geometry  may be attributed to the size-dependent forces  exerted by the planet on the grains.

\subsubsection{Shadows}\label{sec-shadow}
The observed asymmetry of the transition disks in the form of shadows is widely attributed to the inner disk inclined significantly relative to the outer disk, resulting in the casting of varying-depth shadows onto the outer disk \citep{Bohn2022A&A}. 
%With the combination of near infrared and submillimeter interferometric observations of 20 well-known transition disks, \citet{Bohn2022A&A} measured the misalignments between their inner and outer disk structures and successfully explained the shadows of the outer disks.
\citet{Min2017} provided analytical equations describing the positions of these shadow features as a function of the orientation between the inner and outer disk and the height of the outer disk wall, such as HD100453. Based on the hydrodynamic simulations combined with three-dimensional radiative transfer calculations, \citet{Benisty2018A&A} adopted the warped disk model and reproduced all major morphological features for HD 143006. Similarly \citet{MuroArena2020} found that a single misaligned ring was insufficient to account for the wide shadow and instead utilized  two separate misaligned zones to effectively  reproduce most of the observed features within the protoplanetary disk of HD~139614.
Through a comparative analysis of STIS images in 2000 and those from NICMOS in 1998, 2004, and 2005, \citet{Debes2017ApJ} quantified a constant angular velocity in the azimuthal brightness asymmetry of the TW Hya disk, manifesting a counterclockwise direction.
Recently, \citet{Debes2023} reported the presence of two distinct shadows observed in the TW Hya that changed from one in \citet{Debes2017ApJ}.%by employing two misaligned rings with small inclination relative to the outer disk. 

We detected potential  physical substructures exhibiting pixelwise average  S/N exceeding 1, including  two shadows, and one potential inner dust. After subtracting the best-fitting model and the stellar PSF components, we present the residual images, in Fig.~\ref{fig:fig-all}. Specifically, we observed one round shadow and two indistinct shadows in the  $K$-bands on 2015 June 10, respectively denoted as `A' and `B?', `C?' in Fig.~\ref{fig:fig-all}. From the observation in the  $H$-bands on 2015 Aug 13, we identified one narrow shadow accompanied by another wider shadow, respectively denoted as `B', `C' in Fig.~\ref{fig:fig-all}, both of which closely neighbored two indistinct shadows in the $K$-bands. Additionally, the $L'$-band observations on May 10, 2017,  revealed the presence of  two wide shadow and an additional inconspicuous shadow, respectively denoted as `B', `C' and `D?' in Fig.~\ref{fig:fig-all}. To quantify the asymmetry, we computed radially averaged azimuthal profiles of the disk reduced data, using apertures of 6 pixel in radius placed near the position of $r_{\rm c}$ , after deprojecting the image using $\theta_{\rm inc} = 6^\circ$ and $\theta_{\rm PA} = 258\fdg7$ and  scaling by its distance from the central star. The resulting figure is depicted in Fig.~\ref{fig:fig-brightness_R2scale}

The temporal variability of the shadows is in accordance with the findings of \citet{Pinilla2018}. In comparison to the shadow observations reported by \citet{MuroArena2020} and \citet{Debes2023}, the shadows  observed in this study have several  similar dynamics features. 
%Firstly, a single shadow evolves into two wider shadows over time 
Similarly, the number of the shadows exhibits temporal variation. Furthermore, Considering the observed data for J1604, we could infer the existence of potential multiply misaligned and  dynamic rings within the inner region, which may contribute to the generation of wide and variable shadows. This inference suggests that the brighter inner side could coincide with the fainter region of the outer disk. However, unlike the two overlapping shadows reported by \citet{Debes2023}, our $H$-band observations revealed non-overlapping shadows. In contrast to the multiple arcs feature out of the parallactic angle range of the shadow reported by \citet{MuroArena2020}, we detected only one possible arc in the $H3$-band observation.  
Taking in account the observation by \citet{Pinilla2018}, we inferred that the shadow of J1604 occupied in `A' position between 2012 and 2015. We confirmed the change of the shadows number from one to two occurred in 2015. It is likely that `A' shadow disappeared in 2015, and then the shadows in B and C position gradually became evident, which was distinct from the rotation of the shadows and can not be explained by the precession of an inner disk \citep{Debes2017ApJ}. The observations here report a distinct and previously less explored dynamical characteristics of inner disks.

Due to the small dust mass and large inclination, small separation, the inner disk remains inaccessible from direct imaging instruments. We detected positive residuals interior to one shadow (`C') on the outer disk. If the signal was true, it could be residual dust right outside the coronagraph that is better revealed after disk modeling.  The dust can block more light and make the shadow on the outer disk darker and wider than another shadow (`B'). While this may be a spiral running away from the inner disk due to a massive coronal mass ejection (CME) along the equatorial plane of the star \citep[e.g.,][]{Olofsson2018}, it should be noted that CME is for M stars and not applicable for J1604 which is a K2 star.  
%the spiral as the consequence of a massive coronal mass ejection (CME) along the equatorial plane of the star. 
%Conversely, it is also a possibility that the dust particles have migrated away from the outer disk, as described in \citet{Jiang2023}, which detailed the phenomenon of pebbles diffusion from the ring inner edge.

The residual images of other bands also reveal discernible shadows at varying positions, see Fig.~\ref{fig:fig-all}.  Due to the low S/N, we do not discuss more about the residual images in other bands (i.e., $K$, $L'$). 

\section{Conclusion}\label{sec-sum}

We conducted $L'$-band observations of the J1604 protoplanetary disk system using the Keck/NIRC2 vortex coronagraph on UT 2017 May 10. We also retrieved archival SPHERE observations of J1604 on 2015 August 13 in $H2$- and $H3$-band, and on 2015 June 10 in $K1$- and $K2$-band. Using KLIP post-processing with reference differential imaging (either exposures in nearby nights for NIRC2, or archival exposures for IRDIS), we obtained the images of J1604 disk. We forward modeled the disk in different bands to obtain its original surface brightness.

Using forward modeling and comparing with stellar photometry (within $1.22\lambda/D$), we obtained broadband relative reflectance for J1604 in total intensity imaging observations. Together shorter-wavelength SPHERE study in polarized light and for $R$- and $H$-band \citep{2023A&AMa}, our J1604 reflectance measurements span a broad wavelength range from 0.5~$\mu$m to 4~$\mu$m. 
On the one hand, from 0.5~$\mu$m to 2.1~$\mu$m, we observe a monotonically decreasing trend in its reflectance. This trend contrasts with previous work showing gradual increasing trend debris disks (e.g., HD~114082, HD~117214, HR~4796A), illustrating potential mineralogical difference in different evolutionary stages of circumstellar disks. The evolution of the disk might thus have a progressive transition in the reflectance spectrum shape, shifting from J1604-like characteristic blue slope to those featuring red slope resembling HR 4796A. 
On the other hand, from 2.1~$\mu$m to 4$~\mu$m, we observe a relatively flat reflectance. This differs from %\sout{the significant and distinctive variations} 
the phenomenon that the reflectance at ${\sim}3.8 \mu$m is larger than that at ${\sim}2.2\mu$m observed in the reflectance of AB~Aur, HD~142527, and HD~141569, suggesting the difference of dust mineralogy potentially due to different %\sout{in the} 
protoplanetary disk stage or even different molecular cloud origins. Moving forward, to further explore the trends and differences in the reflectance of circumstellar disks, future studies on reflectance spectra with integral field spectroghaphs (e.g., SPHERE, Gemini planet Imager) that can spatially resolve circumstellar systems are necessary.

From the shadow variation that cannot be explained by the precession of the inner disk, we are in the era of witnessing a possible, distinct and previously less explored dynamical characteristics of inner disks in protoplanetary systems \citep[e.g.,][]{Debes2023}. %In the J1604 observations at different wavelengths, our best-fit disk models show different power law profiles for the radial brightness distribution, indicating dust particle condensation at the inner edge of disks. 
We identified a shadow that vanished in 2015 (labeled `A'), and two shadow that potentially enhanced then (labeled `B' and `C'). Our modeling residuals show potential dust components that can cast shadows on outer disk. Future multi-epoch follow-up of outer disks in scattered light could help us probe the dynamics of the inner disks and study the connection of the two regions \citep[e.g.,][]{Bohn2022A&A}.

\begin{acknowledgements}
We thank Myriam Benisty for discussions on shadowing effects, and Juan Quiroz for sharing a disk modeling example. We thank Benoit Carry, Pierre Beck, and Oliver Poch for discusssions on dust reflectance.
%France/EU acknoledgement 
This research has received funding from the European Union's Horizon Europe research and innovation programme under the Marie Sk\l odowska-Curie grant agreement No.~101103114. This project has received funding from the European Research Council (ERC) under the European Union's Horizon 2020 research and innovation programme (PROTOPLANETS, grant agreement No.~101002188). 
%SYSU acknowledgement 
We acknowledge the financial support from the National Key R\&D Program of China (2020YFC2201400), NSFC grant 12073092, 12103097, 12103098, 11733006, the science research grants from the China Manned Space Project (No.~CMS-CSST-2021-B09), Guangdong Major Project of Basic and Applied Basic Research (grant No.~2019B030302001), Guangzhou Basic and Applied Basic Research Program (202102080371), and the Fundamental Research Funds for the Central Universities, Sun Yat-sen University.
%SPHERE data acknowledgement
Based on observations performed with VLT/SPHERE under program ID 095.C-0673(A) and 295.C-5034(A).
%Caltech and Keck acknoledgement
This research is partially supported by NASA ROSES XRP, award 80NSSC19K0294. Some of the data presented herein were obtained at the W.~M.~Keck Observatory, which is operated as a scientific partnership among the California Institute of Technology, the University of California and the National Aeronautics and Space Administration. The Observatory was made possible by the generous financial support of the W.~M.~Keck Foundation. The authors wish to recognize and acknowledge the very significant cultural role and reverence that the summit of Maunakea has always had within the indigenous Hawaiian community.  We are most fortunate to have the opportunity to conduct observations from this mountain. Part of the computations presented here were conducted in the Resnick High Performance Computing Center, a facility supported by Resnick Sustainability Institute at the California Institute of Technology.
\end{acknowledgements}

\bibliography{refs}

\begin{thebibliography}{}
\expandafter\ifx\csname natexlab\endcsname\relax\def\natexlab#1{#1}\fi
\providecommand{\url}[1]{\href{#1}{#1}}
\providecommand{\dodoi}[1]{}
\providecommand{\doarXiv}[1]{\href{https://arxiv.org/abs/#1}{\nolinkurl{https://arxiv.org/abs/#1}}}

\bibitem[{{Adibekyan} {et~al.}(2021){Adibekyan}, {Dorn}, {Sousa}, {Santos},
  {Bitsch}, {Israelian}, {Mordasini}, {Barros}, {Delgado Mena}, {Demangeon},
  {Faria}, {Figueira}, {Hakobyan}, {Oshagh}, {Soares}, {Kunitomo}, {Takeda},
  {Jofr{\'e}}, {Petrucci}, \& {Martioli}}]{Adibekyan2021}
{Adibekyan}, V., {Dorn}, C., {Sousa}, S.~G., {et~al.} 2021,
  \href{http://dx.doi.org/10.1126/science.abg8794}{\color{magenta}Science},
  \href{https://ui.adsabs.harvard.edu/abs/2021Sci...374..330A}{\color{blue}374},
  \href{https://ui.adsabs.harvard.edu/abs/2021Sci...374..330A}{\color{blue}330}

\bibitem[{{Ansdell} {et~al.}(2016){Ansdell}, {Gaidos}, {Williams}, {Kennedy},
  {Wyatt}, {LaCourse}, {Jacobs}, \& {Mann}}]{Ansdell2016MNRAS}
{Ansdell}, M., {Gaidos}, E., {Williams}, J.~P., {et~al.} 2016,
  \href{http://dx.doi.org/10.1093/mnrasl/slw140}{\color{magenta}\mnras},
  \href{https://ui.adsabs.harvard.edu/abs/2016MNRAS.462L.101A}{\color{blue}462},
  \href{https://ui.adsabs.harvard.edu/abs/2016MNRAS.462L.101A}{\color{blue}L101}

\bibitem[{{Augereau} {et~al.}(1999){Augereau}, {Lagrange}, {Mouillet},
  {Papaloizou}, \& {Grorod}}]{Augereau1999}
{Augereau}, J.~C., {Lagrange}, A.~M., {Mouillet}, D., {et~al.} 1999,
  \href{http://dx.doi.org/10.48550/arXiv.astro-ph/9906429}{\color{magenta}\aap},
  \href{https://ui.adsabs.harvard.edu/abs/1999A&A...348..557A}{\color{blue}348},
  \href{https://ui.adsabs.harvard.edu/abs/1999A&A...348..557A}{\color{blue}557}

\bibitem[{{Barenfeld} {et~al.}(2016){Barenfeld}, {Carpenter}, {Ricci}, \&
  {Isella}}]{Barenfeld2016}
{Barenfeld}, S.~A., {Carpenter}, J.~M., {Ricci}, L., \& {Isella}, A. 2016,
  \href{http://dx.doi.org/10.3847/0004-637X/827/2/142}{\color{magenta}\apj},
  \href{https://ui.adsabs.harvard.edu/abs/2016ApJ...827..142B}{\color{blue}827},
  \href{https://ui.adsabs.harvard.edu/abs/2016ApJ...827..142B}{\color{blue}142}

\bibitem[{{Benisty} {et~al.}(2018){Benisty}, {Juh{\'a}sz}, {Facchini},
  {Pinilla}, {de Boer}, {P{\'e}rez}, {Keppler}, {Muro-Arena}, {Villenave},
  {Andrews}, {Dominik}, {Dullemond}, {Gallenne}, {Garufi}, {Ginski}, \&
  {Isella}}]{Benisty2018A&A}
{Benisty}, M., {Juh{\'a}sz}, A., {Facchini}, S., {et~al.} 2018,
  \href{http://dx.doi.org/10.1051/0004-6361/201833913}{\color{magenta}\aap},
  \href{https://ui.adsabs.harvard.edu/abs/2018A&A...619A.171B}{\color{blue}619},
  \href{https://ui.adsabs.harvard.edu/abs/2018A&A...619A.171B}{\color{blue}A171}

\bibitem[{{Betti} {et~al.}(2022){Betti}, {Follette}, {Jorquera}, {Duch{\^e}ne},
  {Mazoyer}, {Bonnefoy}, {Chauvin}, {P{\'e}rez}, {Boccaletti}, {Pinte},
  {Weinberger}, {Grady}, {Close}, {Defr{\`e}re}, {Downey}, {Hinz},
  {M{\'e}nard}, {Schneider}, {Skemer}, \& {Vaz}}]{Betti2022AJ}
{Betti}, S.~K., {Follette}, K., {Jorquera}, S., {et~al.} 2022,
  \href{http://dx.doi.org/10.3847/1538-3881/ac4d9b}{\color{magenta}\aj},
  \href{https://ui.adsabs.harvard.edu/abs/2022AJ....163..145B}{\color{blue}163},
  \href{https://ui.adsabs.harvard.edu/abs/2022AJ....163..145B}{\color{blue}145}

\bibitem[{{Beuzit} {et~al.}(2019){Beuzit}, {Vigan}, {Mouillet}, {Dohlen},
  {Gratton}, {Boccaletti}, {Sauvage}, {Schmid}, {Langlois}, {Petit},
  {Baruffolo}, {Feldt}, {Milli}, {Wahhaj}, {Abe}, {Anselmi}, {Antichi},
  {Barette}, {Baudrand}, {Baudoz}, {Bazzon}, {Bernardi}, {Blanchard}, {Brast},
  {Bruno}, {Buey}, {Carbillet}, {Carle}, {Cascone}, {Chapron}, {Charton},
  {Chauvin}, {Claudi}, {Costille}, {De Caprio}, {de Boer}, {Delboulb{\'e}},
  {Desidera}, {Dominik}, {Downing}, {Dupuis}, {Fabron}, {Fantinel}, {Farisato},
  {Feautrier}, {Fedrigo}, {Fusco}, {Gigan}, {Ginski}, {Girard}, {Giro},
  {Gisler}, {Gluck}, {Gry}, {Henning}, {Hubin}, {Hugot}, {Incorvaia}, {Jaquet},
  {Kasper}, {Lagadec}, {Lagrange}, {Le Coroller}, {Le Mignant}, {Le Ruyet},
  {Lessio}, {Lizon}, {Llored}, {Lundin}, {Madec}, {Magnard}, {Marteaud},
  {Martinez}, {Maurel}, {M{\'e}nard}, {Mesa}, {M{\"o}ller-Nilsson}, {Moulin},
  {Moutou}, {Orign{\'e}}, {Parisot}, {Pavlov}, {Perret}, {Pragt}, {Puget},
  {Rabou}, {Ramos}, {Reess}, {Rigal}, {Rochat}, {Roelfsema}, {Rousset}, {Roux},
  {Saisse}, {Salasnich}, {Santambrogio}, {Scuderi}, {Segransan}, {Sevin},
  {Siebenmorgen}, {Soenke}, {Stadler}, {Suarez}, {Tiph{\`e}ne}, {Turatto},
  {Udry}, {Vakili}, {Waters}, {Weber}, {Wildi}, {Zins}, \&
  {Zurlo}}]{Beuzit2019}
{Beuzit}, J.~L., {Vigan}, A., {Mouillet}, D., {et~al.} 2019,
  \href{http://dx.doi.org/10.1051/0004-6361/201935251}{\color{magenta}\aap},
  \href{https://ui.adsabs.harvard.edu/abs/2019A&A...631A.155B}{\color{blue}631},
  \href{https://ui.adsabs.harvard.edu/abs/2019A&A...631A.155B}{\color{blue}A155}

\bibitem[{{Bhowmik} {et~al.}(2019){Bhowmik}, {Boccaletti}, {Th{\'e}bault},
  {Kral}, {Mazoyer}, {Milli}, {Maire}, {van Holstein}, {Augereau}, {Baudoz},
  {Feldt}, {Galicher}, {Henning}, {Lagrange}, {Olofsson}, {Pantin}, \&
  {Perrot}}]{Bhowmik2019}
{Bhowmik}, T., {Boccaletti}, A., {Th{\'e}bault}, P., {et~al.} 2019,
  \href{http://dx.doi.org/10.1051/0004-6361/201936076}{\color{magenta}\aap},
  \href{https://ui.adsabs.harvard.edu/abs/2019A&A...630A..85B}{\color{blue}630},
  \href{https://ui.adsabs.harvard.edu/abs/2019A&A...630A..85B}{\color{blue}A85}

\bibitem[{{Bohn} {et~al.}(2022){Bohn}, {Benisty}, {Perraut}, {van der Marel},
  {W{\"o}lfer}, {van Dishoeck}, {Facchini}, {Manara}, {Teague}, {Francis},
  {Berger}, {Garcia-Lopez}, {Ginski}, {Henning}, {Kenworthy}, {Kraus},
  {M{\'e}nard}, {M{\'e}rand}, \& {P{\'e}rez}}]{Bohn2022A&A}
{Bohn}, A.~J., {Benisty}, M., {Perraut}, K., {et~al.} 2022,
  \href{http://dx.doi.org/10.1051/0004-6361/202142070}{\color{magenta}\aap},
  \href{https://ui.adsabs.harvard.edu/abs/2022A&A...658A.183B}{\color{blue}658},
  \href{https://ui.adsabs.harvard.edu/abs/2022A&A...658A.183B}{\color{blue}A183}

\bibitem[{{Canovas} {et~al.}(2017){Canovas}, {Hardy}, {Zurlo}, {Wahhaj},
  {Schreiber}, {Vigan}, {Villaver}, {Olofsson}, {Meeus}, {M{\'e}nard},
  {Caceres}, {Cieza}, \& {Garufi}}]{Canovas2017}
{Canovas}, H., {Hardy}, A., {Zurlo}, A., {et~al.} 2017,
  \href{http://dx.doi.org/10.1051/0004-6361/201629145}{\color{magenta}\aap},
  \href{https://ui.adsabs.harvard.edu/abs/2017A&A...598A..43C}{\color{blue}598},
  \href{https://ui.adsabs.harvard.edu/abs/2017A&A...598A..43C}{\color{blue}A43}

\bibitem[{{Davies}(2019)}]{Davies2019}
{Davies}, C.~L. 2019,
  \href{http://dx.doi.org/10.1093/mnras/stz086}{\color{magenta}\mnras},
  \href{https://ui.adsabs.harvard.edu/abs/2019MNRAS.484.1926D}{\color{blue}484},
  \href{https://ui.adsabs.harvard.edu/abs/2019MNRAS.484.1926D}{\color{blue}1926}

\bibitem[{{Debes} {et~al.}(2023){Debes}, {Nealon}, {Alexander}, {Weinberger},
  {Wolff}, {Hines}, {Kastner}, {Jang-Condell}, {Pinte}, {Plavchan}, \&
  {Pueyo}}]{Debes2023}
{Debes}, J., {Nealon}, R., {Alexander}, R., {et~al.} 2023,
  \href{http://dx.doi.org/10.3847/1538-4357/acbdf1}{\color{magenta}\apj},
  \href{https://ui.adsabs.harvard.edu/abs/2023ApJ...948...36D}{\color{blue}948},
  \href{https://ui.adsabs.harvard.edu/abs/2023ApJ...948...36D}{\color{blue}36}

\bibitem[{{Debes} {et~al.}(2008){Debes}, {Weinberger}, \&
  {Schneider}}]{Debes2008}
{Debes}, J.~H., {Weinberger}, A.~J., \& {Schneider}, G. 2008,
  \href{http://dx.doi.org/10.1086/527546}{\color{magenta}\apjl},
  \href{https://ui.adsabs.harvard.edu/abs/2008ApJ...673L.191D}{\color{blue}673},
  \href{https://ui.adsabs.harvard.edu/abs/2008ApJ...673L.191D}{\color{blue}L191}

\bibitem[{{Debes} {et~al.}(2017){Debes}, {Poteet}, {Jang-Condell}, {Gaspar},
  {Hines}, {Kastner}, {Pueyo}, {Rapson}, {Roberge}, {Schneider}, \&
  {Weinberger}}]{Debes2017ApJ}
{Debes}, J.~H., {Poteet}, C.~A., {Jang-Condell}, H., {et~al.} 2017,
  \href{http://dx.doi.org/10.3847/1538-4357/835/2/205}{\color{magenta}\apj},
  \href{https://ui.adsabs.harvard.edu/abs/2017ApJ...835..205D}{\color{blue}835},
  \href{https://ui.adsabs.harvard.edu/abs/2017ApJ...835..205D}{\color{blue}205}

\bibitem[{{Dohlen} {et~al.}(2008){Dohlen}, {Langlois}, {Saisse}, {Hill},
  {Origne}, {Jacquet}, {Fabron}, {Blanc}, {Llored}, {Carle}, {Moutou}, {Vigan},
  {Boccaletti}, {Carbillet}, {Mouillet}, \& {Beuzit}}]{IRDIS_Dohlen2008SPIE}
{Dohlen}, K., {Langlois}, M., {Saisse}, M., {et~al.} 2008,
  \href{http://dx.doi.org/10.1117/12.789786}{\color{magenta}Proc.~SPIE},
  \href{https://ui.adsabs.harvard.edu/abs/2008SPIE.7014E..3LD}{\color{blue}7014},
  \href{https://ui.adsabs.harvard.edu/abs/2008SPIE.7014E..3LD}{\color{blue}70143L}

\bibitem[{{Dong} {et~al.}(2017){Dong}, {van der Marel}, {Hashimoto}, {Chiang},
  {Akiyama}, {Liu}, {Muto}, {Knapp}, {Tsukagoshi}, {Brown}, {Bruderer},
  {Koyamatsu}, {Kudo}, {Ohashi}, {Rich}, {Satoshi}, {Takami}, {Wisniewski},
  {Yang}, {Zhu}, \& {Tamura}}]{Dong2017}
{Dong}, R., {van der Marel}, N., {Hashimoto}, J., {et~al.} 2017,
  \href{http://dx.doi.org/10.3847/1538-4357/aa5abf}{\color{magenta}\apj},
  \href{https://ui.adsabs.harvard.edu/abs/2017ApJ...836..201D}{\color{blue}836},
  \href{https://ui.adsabs.harvard.edu/abs/2017ApJ...836..201D}{\color{blue}201}

\bibitem[{{Dressing} {et~al.}(2015){Dressing}, {Charbonneau}, {Dumusque},
  {Gettel}, {Pepe}, {Collier Cameron}, {Latham}, {Molinari}, {Udry}, {Affer},
  {Bonomo}, {Buchhave}, {Cosentino}, {Figueira}, {Fiorenzano}, {Harutyunyan},
  {Haywood}, {Johnson}, {Lopez-Morales}, {Lovis}, {Malavolta}, {Mayor},
  {Micela}, {Motalebi}, {Nascimbeni}, {Phillips}, {Piotto}, {Pollacco},
  {Queloz}, {Rice}, {Sasselov}, {S{\'e}gransan}, {Sozzetti}, {Szentgyorgyi}, \&
  {Watson}}]{Dressing2015ApJ}
{Dressing}, C.~D., {Charbonneau}, D., {Dumusque}, X., {et~al.} 2015,
  \href{http://dx.doi.org/10.1088/0004-637X/800/2/135}{\color{magenta}\apj},
  \href{https://ui.adsabs.harvard.edu/abs/2015ApJ...800..135D}{\color{blue}800},
  \href{https://ui.adsabs.harvard.edu/abs/2015ApJ...800..135D}{\color{blue}135}

\bibitem[{{Engler} {et~al.}(2023){Engler}, {Milli}, {Gratton}, {Ulmer-Moll},
  {Vigan}, {Lagrange}, {Kiefer}, {Rubini}, {Grandjean}, {Schmid}, {Messina},
  {Squicciarini}, {Olofsson}, {Th{\'e}bault}, {van Holstein}, {Janson},
  {M{\'e}nard}, {Marshall}, {Chauvin}, {Lendl}, {Bhowmik}, {Boccaletti},
  {Bonnefoy}, {del Burgo}, {Choquet}, {Desidera}, {Feldt}, {Fusco}, {Girard},
  {Gisler}, {Hagelberg}, {Langlois}, {Maire}, {Mesa}, {Meyer}, {Rabou},
  {Rodet}, {Schmidt}, \& {Zurlo}}]{Engler2023A&A}
{Engler}, N., {Milli}, J., {Gratton}, R., {et~al.} 2023,
  \href{http://dx.doi.org/10.1051/0004-6361/202244380}{\color{magenta}\aap},
  \href{https://ui.adsabs.harvard.edu/abs/2023A&A...672A...1E}{\color{blue}672},
  \href{https://ui.adsabs.harvard.edu/abs/2023A&A...672A...1E}{\color{blue}A1}

\bibitem[{{Foreman-Mackey} {et~al.}(2013){Foreman-Mackey}, {Hogg}, {Lang}, \&
  {Goodman}}]{Foreman2013}
{Foreman-Mackey}, D., {Hogg}, D.~W., {Lang}, D., \& {Goodman}, J. 2013,
  \href{http://dx.doi.org/10.1086/670067}{\color{magenta}\pasp},
  \href{https://ui.adsabs.harvard.edu/abs/2013PASP..125..306F}{\color{blue}125},
  \href{https://ui.adsabs.harvard.edu/abs/2013PASP..125..306F}{\color{blue}306}

\bibitem[{{Fouesneau} {et~al.}(2022){Fouesneau}, {Andrae}, {Dharmawardena},
  {Rybizki}, {Bailer-Jones}, \& {Demleitner}}]{Fouesneau2022A&A}
{Fouesneau}, M., {Andrae}, R., {Dharmawardena}, T., {et~al.} 2022,
  \href{http://dx.doi.org/10.1051/0004-6361/202141828}{\color{magenta}\aap},
  \href{https://ui.adsabs.harvard.edu/abs/2022A&A...662A.125F}{\color{blue}662},
  \href{https://ui.adsabs.harvard.edu/abs/2022A&A...662A.125F}{\color{blue}A125}

\bibitem[{{Francis} \& {van der Marel}(2020)}]{Francis2020}
{Francis}, L., \& {van der Marel}, N. 2020,
  \href{http://dx.doi.org/10.3847/1538-4357/ab7b63}{\color{magenta}\apj},
  \href{https://ui.adsabs.harvard.edu/abs/2020ApJ...892..111F}{\color{blue}892},
  \href{https://ui.adsabs.harvard.edu/abs/2020ApJ...892..111F}{\color{blue}111}

\bibitem[{{Gaia Collaboration} {et~al.}(2023){Gaia Collaboration}, {Vallenari},
  {Brown}, {Prusti}, {de Bruijne}, {Arenou}, {Babusiaux}, {Biermann},
  {Creevey}, {Ducourant}, {Evans}, {Eyer}, {Guerra}, {Hutton}, {Jordi},
  {Klioner}, {Lammers}, {Lindegren}, {Luri}, {Mignard}, {Panem}, {Pourbaix},
  {Randich}, {Sartoretti}, {Soubiran}, {Tanga}, {Walton}, {Bailer-Jones},
  {Bastian}, {Drimmel}, {Jansen}, {Katz}, {Lattanzi}, {van Leeuwen}, {Bakker},
  {Cacciari}, {Casta{\~n}eda}, {De Angeli}, {Fabricius}, {Fouesneau},
  {Fr{\'e}mat}, {Galluccio}, {Guerrier}, {Heiter}, {Masana}, {Messineo},
  {Mowlavi}, {Nicolas}, {Nienartowicz}, {Pailler}, {Panuzzo}, {Riclet}, {Roux},
  {Seabroke}, {Sordo}, {Th{\'e}venin}, {Gracia-Abril}, {Portell}, {Teyssier},
  {Altmann}, {Andrae}, {Audard}, {Bellas-Velidis}, {Benson}, {Berthier},
  {Blomme}, {Burgess}, {Busonero}, {Busso}, {C{\'a}novas}, {Carry}, {Cellino},
  {Cheek}, {Clementini}, {Damerdji}, {Davidson}, {de Teodoro}, {Nu{\~n}ez
  Campos}, {Delchambre}, {Dell'Oro}, {Esquej}, {Fern{\'a}ndez-Hern{\'a}ndez},
  {Fraile}, {Garabato}, {Garc{\'\i}a-Lario}, {Gosset}, {Haigron}, {Halbwachs},
  {Hambly}, {Harrison}, {Hern{\'a}ndez}, {Hestroffer}, {Hodgkin}, {Holl},
  {Jan{\ss}en}, {Jevardat de Fombelle}, {Jordan}, {Krone-Martins}, {Lanzafame},
  {L{\"o}ffler}, {Marchal}, {Marrese}, {Moitinho}, {Muinonen}, {Osborne},
  {Pancino}, {Pauwels}, {Recio-Blanco}, {Reyl{\'e}}, {Riello}, {Rimoldini},
  {Roegiers}, {Rybizki}, {Sarro}, {Siopis}, {Smith}, {Sozzetti}, {Utrilla},
  {van Leeuwen}, {Abbas}, {{\'A}brah{\'a}m}, {Abreu Aramburu}, {Aerts},
  {Aguado}, {Ajaj}, {Aldea-Montero}, {Altavilla}, {{\'A}lvarez}, {Alves},
  {Anders}, {Anderson}, {Anglada Varela}, {Antoja}, {Baines}, {Baker},
  {Balaguer-N{\'u}{\~n}ez}, {Balbinot}, {Balog}, {Barache}, {Barbato},
  {Barros}, {Barstow}, {Bartolom{\'e}}, {Bassilana}, {Bauchet}, {Becciani},
  {Bellazzini}, {Berihuete}, {Bernet}, {Bertone}, {Bianchi}, {Binnenfeld},
  {Blanco-Cuaresma}, {Blazere}, {Boch}, {Bombrun}, {Bossini}, {Bouquillon},
  {Bragaglia}, {Bramante}, {Breedt}, {Bressan}, {Brouillet}, {Brugaletta},
  {Bucciarelli}, {Burlacu}, {Butkevich}, {Buzzi}, {Caffau}, {Cancelliere},
  {Cantat-Gaudin}, {Carballo}, {Carlucci}, {Carnerero}, {Carrasco},
  {Casamiquela}, {Castellani}, {Castro-Ginard}, {Chaoul}, {Charlot}, {Chemin},
  {Chiaramida}, {Chiavassa}, {Chornay}, {Comoretto}, {Contursi}, {Cooper},
  {Cornez}, {Cowell}, {Crifo}, {Cropper}, {Crosta}, {Crowley}, {Dafonte},
  {Dapergolas}, {David}, {David}, {de Laverny}, {De Luise}, {De March}, {De
  Ridder}, {de Souza}, {de Torres}, {del Peloso}, {del Pozo}, {Delbo},
  {Delgado}, {Delisle}, {Demouchy}, {Dharmawardena}, {Di Matteo}, {Diakite},
  {Diener}, {Distefano}, {Dolding}, {Edvardsson}, {Enke}, {Fabre}, {Fabrizio},
  {Faigler}, {Fedorets}, {Fernique}, {Fienga}, {Figueras}, {Fournier},
  {Fouron}, {Fragkoudi}, {Gai}, {Garcia-Gutierrez}, {Garcia-Reinaldos},
  {Garc{\'\i}a-Torres}, {Garofalo}, {Gavel}, {Gavras}, {Gerlach}, {Geyer},
  {Giacobbe}, {Gilmore}, {Girona}, {Giuffrida}, {Gomel}, {Gomez},
  {Gonz{\'a}lez-N{\'u}{\~n}ez}, {Gonz{\'a}lez-Santamar{\'\i}a},
  {Gonz{\'a}lez-Vidal}, {Granvik}, {Guillout}, {Guiraud},
  {Guti{\'e}rrez-S{\'a}nchez}, {Guy}, {Hatzidimitriou}, {Hauser}, {Haywood},
  {Helmer}, {Helmi}, {Sarmiento}, {Hidalgo}, {Hilger}, {H{\l}adczuk}, {Hobbs},
  {Holland}, {Huckle}, {Jardine}, {Jasniewicz}, {Jean-Antoine Piccolo},
  {Jim{\'e}nez-Arranz}, {Jorissen}, {Juaristi Campillo}, {Julbe}, {Karbevska},
  {Kervella}, {Khanna}, {Kontizas}, {Kordopatis}, {Korn}, {K{\'o}sp{\'a}l},
  {Kostrzewa-Rutkowska}, {Kruszy{\'n}ska}, {Kun}, {Laizeau}, {Lambert},
  {Lanza}, {Lasne}, {Le Campion}, {Lebreton}, {Lebzelter}, {Leccia}, {Leclerc},
  {Lecoeur-Taibi}, {Liao}, {Licata}, {Lindstr{\o}m}, {Lister}, {Livanou},
  {Lobel}, {Lorca}, {Loup}, {Madrero Pardo}, {Magdaleno Romeo}, {Managau},
  {Mann}, {Manteiga}, {Marchant}, {Marconi}, {Marcos}, {Marcos Santos},
  {Mar{\'\i}n Pina}, {Marinoni}, {Marocco}, {Marshall}, {Martin Polo},
  {Mart{\'\i}n-Fleitas}, {Marton}, {Mary}, {Masip}, {Massari},
  {Mastrobuono-Battisti}, {Mazeh}, {McMillan}, {Messina}, {Michalik}, {Millar},
  {Mints}, {Molina}, {Molinaro}, {Moln{\'a}r}, {Monari}, {Mongui{\'o}},
  {Montegriffo}, {Montero}, {Mor}, {Mora}, {Morbidelli}, {Morel}, {Morris},
  {Muraveva}, {Murphy}, {Musella}, {Nagy}, {Noval}, {Oca{\~n}a}, {Ogden},
  {Ordenovic}, {Osinde}, {Pagani}, {Pagano}, {Palaversa}, {Palicio},
  {Pallas-Quintela}, {Panahi}, {Payne-Wardenaar}, {Pe{\~n}alosa Esteller},
  {Penttil{\"a}}, {Pichon}, {Piersimoni}, {Pineau}, {Plachy}, {Plum}, {Poggio},
  {Pr{\v{s}}a}, {Pulone}, {Racero}, {Ragaini}, {Rainer}, {Raiteri}, {Rambaux},
  {Ramos}, {Ramos-Lerate}, {Re Fiorentin}, {Regibo}, {Richards}, {Rios Diaz},
  {Ripepi}, {Riva}, {Rix}, {Rixon}, {Robichon}, {Robin}, {Robin}, {Roelens},
  {Rogues}, {Rohrbasser}, {Romero-G{\'o}mez}, {Rowell}, {Royer}, {Ruz Mieres},
  {Rybicki}, {Sadowski}, {S{\'a}ez N{\'u}{\~n}ez}, {Sagrist{\`a} Sell{\'e}s},
  {Sahlmann}, {Salguero}, {Samaras}, {Sanchez Gimenez}, {Sanna},
  {Santove{\~n}a}, {Sarasso}, {Schultheis}, {Sciacca}, {Segol}, {Segovia},
  {S{\'e}gransan}, {Semeux}, {Shahaf}, {Siddiqui}, {Siebert}, {Siltala},
  {Silvelo}, {Slezak}, {Slezak}, {Smart}, {Snaith}, {Solano}, {Solitro},
  {Souami}, {Souchay}, {Spagna}, {Spina}, {Spoto}, {Steele},
  {Steidelm{\"u}ller}, {Stephenson}, {S{\"u}veges}, {Surdej}, {Szabados},
  {Szegedi-Elek}, {Taris}, {Taylor}, {Teixeira}, {Tolomei}, {Tonello}, {Torra},
  {Torra}, {Torralba Elipe}, {Trabucchi}, {Tsounis}, {Turon}, {Ulla}, {Unger},
  {Vaillant}, {van Dillen}, {van Reeven}, {Vanel}, {Vecchiato}, {Viala},
  {Vicente}, {Voutsinas}, {Weiler}, {Wevers}, {Wyrzykowski}, {Yoldas}, {Yvard},
  {Zhao}, {Zorec}, {Zucker}, \& {Zwitter}}]{Gaia2023A&A}
{Gaia Collaboration}, {Vallenari}, A., {Brown}, A.~G.~A., {et~al.} 2023,
  \href{http://dx.doi.org/10.1051/0004-6361/202243940}{\color{magenta}\aap},
  \href{https://ui.adsabs.harvard.edu/abs/2023A&A...674A...1G}{\color{blue}674},
  \href{https://ui.adsabs.harvard.edu/abs/2023A&A...674A...1G}{\color{blue}A1}

\bibitem[{{Henyey} \& {Greenstein}(1941)}]{hg41}
{Henyey}, L.~G., \& {Greenstein}, J.~L. 1941,
  \href{http://dx.doi.org/10.1086/144246}{\color{magenta}\apj},
  \href{https://ui.adsabs.harvard.edu/abs/1941ApJ....93...70H}{\color{blue}93},
  \href{https://ui.adsabs.harvard.edu/abs/1941ApJ....93...70H}{\color{blue}70}

\bibitem[{{Honda} {et~al.}(2009){Honda}, {Inoue}, {Fukagawa}, {Oka},
  {Nakamoto}, {Ishii}, {Terada}, {Takato}, {Kawakita}, {Okamoto}, {Shibai},
  {Tamura}, {Kudo}, \& {Itoh}}]{Honda2009ApJ}
{Honda}, M., {Inoue}, A.~K., {Fukagawa}, M., {et~al.} 2009,
  \href{http://dx.doi.org/10.1088/0004-637X/690/2/L110}{\color{magenta}\apjl},
  \href{https://ui.adsabs.harvard.edu/abs/2009ApJ...690L.110H}{\color{blue}690},
  \href{https://ui.adsabs.harvard.edu/abs/2009ApJ...690L.110H}{\color{blue}L110}

\bibitem[{{Honda} {et~al.}(2016){Honda}, {Kudo}, {Takatsuki}, {Inoue},
  {Nakamoto}, {Fukagawa}, {Tamura}, {Terada}, \& {Takato}}]{Honda2016ApJ}
{Honda}, M., {Kudo}, T., {Takatsuki}, S., {et~al.} 2016,
  \href{http://dx.doi.org/10.3847/0004-637X/821/1/2}{\color{magenta}\apj},
  \href{https://ui.adsabs.harvard.edu/abs/2016ApJ...821....2H}{\color{blue}821},
  \href{https://ui.adsabs.harvard.edu/abs/2016ApJ...821....2H}{\color{blue}2}

\bibitem[{{Inoue} {et~al.}(2008){Inoue}, {Honda}, {Nakamoto}, \&
  {Oka}}]{Inoue2008PASJ}
{Inoue}, A.~K., {Honda}, M., {Nakamoto}, T., \& {Oka}, A. 2008,
  \href{http://dx.doi.org/10.1093/pasj/60.3.557}{\color{magenta}\pasj},
  \href{https://ui.adsabs.harvard.edu/abs/2008PASJ...60..557I}{\color{blue}60},
  \href{https://ui.adsabs.harvard.edu/abs/2008PASJ...60..557I}{\color{blue}557}

\bibitem[{{Jiang} \& {Ormel}(2023)}]{Jiang2023}
{Jiang}, H., \& {Ormel}, C.~W. 2023,
  \href{http://dx.doi.org/10.1093/mnras/stac3275}{\color{magenta}\mnras},
  \href{https://ui.adsabs.harvard.edu/abs/2023MNRAS.518.3877J}{\color{blue}518},
  \href{https://ui.adsabs.harvard.edu/abs/2023MNRAS.518.3877J}{\color{blue}3877}

\bibitem[{{Johansen} {et~al.}(2019){Johansen}, {Ida}, \&
  {Brasser}}]{Johansen2019}
{Johansen}, A., {Ida}, S., \& {Brasser}, R. 2019,
  \href{http://dx.doi.org/10.1051/0004-6361/201834071}{\color{magenta}\aap},
  \href{https://ui.adsabs.harvard.edu/abs/2019A&A...622A.202J}{\color{blue}622},
  \href{https://ui.adsabs.harvard.edu/abs/2019A&A...622A.202J}{\color{blue}A202}

\bibitem[{{K{\"o}hler} {et~al.}(2000){K{\"o}hler}, {Kunkel}, {Leinert}, \&
  {Zinnecker}}]{hler2000}
{K{\"o}hler}, R., {Kunkel}, M., {Leinert}, C., \& {Zinnecker}, H. 2000, \aap,
  \href{https://ui.adsabs.harvard.edu/abs/2000A&A...356..541K}{\color{blue}356},
  \href{https://ui.adsabs.harvard.edu/abs/2000A&A...356..541K}{\color{blue}541}

\bibitem[{{Kueny} {et~al.}(2024){Kueny}, {Weinberger}, {Males}, {Morzinski},
  {Close}, {Follette}, \& {Hinz}}]{Kueny2024}
{Kueny}, J.~K., {Weinberger}, A.~J., {Males}, J.~R., {et~al.} 2024,
  \href{http://dx.doi.org/10.3847/1538-4357/ad0f96}{\color{magenta}\apj},
  \href{https://ui.adsabs.harvard.edu/abs/2024ApJ...961...77K}{\color{blue}961},
  \href{https://ui.adsabs.harvard.edu/abs/2024ApJ...961...77K}{\color{blue}77}

\bibitem[{{Ma} \& {Schmid}(2022)}]{Ma2022A&A}
{Ma}, J., \& {Schmid}, H.~M. 2022,
  \href{http://dx.doi.org/10.1051/0004-6361/202142954}{\color{magenta}\aap},
  \href{https://ui.adsabs.harvard.edu/abs/2022A&A...663A.110M}{\color{blue}663},
  \href{https://ui.adsabs.harvard.edu/abs/2022A&A...663A.110M}{\color{blue}A110}

\bibitem[{{Ma} {et~al.}(2023){Ma}, {Schmid}, \& {Tschudi}}]{2023A&AMa}
{Ma}, J., {Schmid}, H.~M., \& {Tschudi}, C. 2023,
  \href{http://dx.doi.org/10.1051/0004-6361/202245697}{\color{magenta}\aap},
  \href{https://ui.adsabs.harvard.edu/abs/2023A&A...676A...6M}{\color{blue}676},
  \href{https://ui.adsabs.harvard.edu/abs/2023A&A...676A...6M}{\color{blue}A6}

\bibitem[{{Mah} {et~al.}(2023){Mah}, {Bitsch}, {Pascucci}, \&
  {Henning}}]{Mah2023A&A}
{Mah}, J., {Bitsch}, B., {Pascucci}, I., \& {Henning}, T. 2023,
  \href{http://dx.doi.org/10.1051/0004-6361/202347169}{\color{magenta}\aap},
  \href{https://ui.adsabs.harvard.edu/abs/2023A&A...677L...7M}{\color{blue}677},
  \href{https://ui.adsabs.harvard.edu/abs/2023A&A...677L...7M}{\color{blue}L7}

\bibitem[{{Maire} {et~al.}(2016){Maire}, {Langlois}, {Dohlen}, {Lagrange},
  {Gratton}, {Chauvin}, {Desidera}, {Girard}, {Milli}, {Vigan}, {Zins},
  {Delorme}, {Beuzit}, {Claudi}, {Feldt}, {Mouillet}, {Puget}, {Turatto}, \&
  {Wildi}}]{Maire2016}
{Maire}, A.-L., {Langlois}, M., {Dohlen}, K., {et~al.} 2016,
  \href{http://dx.doi.org/10.1117/12.2233013}{\color{magenta}Proc.~SPIE},
  \href{https://ui.adsabs.harvard.edu/abs/2016SPIE.9908E..34M}{\color{blue}9908},
  \href{https://ui.adsabs.harvard.edu/abs/2016SPIE.9908E..34M}{\color{blue}990834}

\bibitem[{{Mawet} {et~al.}(2019){Mawet}, {Hirsch}, {Lee}, {Ruffio}, {Bottom},
  {Fulton}, {Absil}, {Beichman}, {Bowler}, {Bryan}, {Choquet}, {Ciardi},
  {Christiaens}, {Defr{\`e}re}, {Gomez Gonzalez}, {Howard}, {Huby}, {Isaacson},
  {Jensen-Clem}, {Kosiarek}, {Marcy}, {Meshkat}, {Petigura}, {Reggiani},
  {Ruane}, {Serabyn}, {Sinukoff}, {Wang}, {Weiss}, \& {Ygouf}}]{Mawet2019}
{Mawet}, D., {Hirsch}, L., {Lee}, E.~J., {et~al.} 2019,
  \href{http://dx.doi.org/10.3847/1538-3881/aaef8a}{\color{magenta}\aj},
  \href{https://ui.adsabs.harvard.edu/abs/2019AJ....157...33M}{\color{blue}157},
  \href{https://ui.adsabs.harvard.edu/abs/2019AJ....157...33M}{\color{blue}33}

\bibitem[{{Mayama} {et~al.}(2018){Mayama}, {Akiyama}, {Pani{\'c}}, {Miley},
  {Tsukagoshi}, {Muto}, {Dong}, {de Leon}, {Mizuki}, {Oh}, {Hashimoto}, {Sai},
  {Currie}, {Takami}, {Grady}, {Hayashi}, {Tamura}, \& {Inutsuka}}]{Mayama2018}
{Mayama}, S., {Akiyama}, E., {Pani{\'c}}, O., {et~al.} 2018,
  \href{http://dx.doi.org/10.3847/2041-8213/aae88b}{\color{magenta}\apjl},
  \href{https://ui.adsabs.harvard.edu/abs/2018ApJ...868L...3M}{\color{blue}868},
  \href{https://ui.adsabs.harvard.edu/abs/2018ApJ...868L...3M}{\color{blue}L3}

\bibitem[{{Mazoyer} {et~al.}(2020){Mazoyer}, {Arriaga}, {Hom},
  {Millar-Blanchaer}, {Chen}, {Wang}, {Duch{\^e}ne}, {Patience}, \&
  {Pueyo}}]{mazoyer20}
{Mazoyer}, J., {Arriaga}, P., {Hom}, J., {et~al.} 2020,
  \href{http://dx.doi.org/10.1117/12.2560091}{\color{magenta}Proc.~SPIE},
  \href{https://ui.adsabs.harvard.edu/abs/2020SPIE11447E..59M}{\color{blue}11447},
  \href{https://ui.adsabs.harvard.edu/abs/2020SPIE11447E..59M}{\color{blue}1144759}

\bibitem[{{Millar-Blanchaer} {et~al.}(2015){Millar-Blanchaer}, {Graham},
  {Pueyo}, {Kalas}, {Dawson}, {Wang}, {Perrin}, {moon}, {Macintosh}, {Ammons},
  {Barman}, {Cardwell}, {Chen}, {Chiang}, {Chilcote}, {Cotten}, {De Rosa},
  {Draper}, {Dunn}, {Duch{\^e}ne}, {Esposito}, {Fitzgerald}, {Follette},
  {Goodsell}, {Greenbaum}, {Hartung}, {Hibon}, {Hinkley}, {Ingraham},
  {Jensen-Clem}, {Konopacky}, {Larkin}, {Long}, {Maire}, {Marchis}, {Marley},
  {Marois}, {Morzinski}, {Nielsen}, {Palmer}, {Oppenheimer}, {Poyneer},
  {Rajan}, {Rantakyr{\"o}}, {Ruffio}, {Sadakuni}, {Saddlemyer}, {Schneider},
  {Sivaramakrishnan}, {Soummer}, {Thomas}, {Vasisht}, {Vega}, {Wallace},
  {Ward-Duong}, {Wiktorowicz}, \& {Wolff}}]{Millar2015ApJ}
{Millar-Blanchaer}, M.~A., {Graham}, J.~R., {Pueyo}, L., {et~al.} 2015,
  \href{http://dx.doi.org/10.1088/0004-637X/811/1/18}{\color{magenta}\apj},
  \href{https://ui.adsabs.harvard.edu/abs/2015ApJ...811...18M}{\color{blue}811},
  \href{https://ui.adsabs.harvard.edu/abs/2015ApJ...811...18M}{\color{blue}18}

\bibitem[{{Miller} \& {Fortney}(2011)}]{Miller2011ApJ}
{Miller}, N., \& {Fortney}, J.~J. 2011,
  \href{http://dx.doi.org/10.1088/2041-8205/736/2/L29}{\color{magenta}\apjl},
  \href{https://ui.adsabs.harvard.edu/abs/2011ApJ...736L..29M}{\color{blue}736},
  \href{https://ui.adsabs.harvard.edu/abs/2011ApJ...736L..29M}{\color{blue}L29}

\bibitem[{{Milli} {et~al.}(2017){Milli}, {Vigan}, {Mouillet}, {Lagrange},
  {Augereau}, {Pinte}, {Mawet}, {Schmid}, {Boccaletti}, {Matr{\`a}}, {Kral},
  {Ertel}, {Chauvin}, {Bazzon}, {M{\'e}nard}, {Beuzit}, {Thalmann}, {Dominik},
  {Feldt}, {Henning}, {Min}, {Girard}, {Galicher}, {Bonnefoy}, {Fusco}, {de
  Boer}, {Janson}, {Maire}, {Mesa}, {Schlieder}, \& {SPHERE
  Consortium}}]{Milli_2017}
{Milli}, J., {Vigan}, A., {Mouillet}, D., {et~al.} 2017,
  \href{http://dx.doi.org/10.1051/0004-6361/201527838}{\color{magenta}\aap},
  \href{https://ui.adsabs.harvard.edu/abs/2017A&A...599A.108M}{\color{blue}599},
  \href{https://ui.adsabs.harvard.edu/abs/2017A&A...599A.108M}{\color{blue}A108}

\bibitem[{{Min} {et~al.}(2017){Min}, {Stolker}, {Dominik}, \&
  {Benisty}}]{Min2017}
{Min}, M., {Stolker}, T., {Dominik}, C., \& {Benisty}, M. 2017,
  \href{http://dx.doi.org/10.1051/0004-6361/201730949}{\color{magenta}\aap},
  \href{https://ui.adsabs.harvard.edu/abs/2017A&A...604L..10M}{\color{blue}604},
  \href{https://ui.adsabs.harvard.edu/abs/2017A&A...604L..10M}{\color{blue}L10}

\bibitem[{{Morbidelli} {et~al.}(2016){Morbidelli}, {Bitsch}, {Crida},
  {Gounelle}, {Guillot}, {Jacobson}, {Johansen}, {Lambrechts}, \&
  {Lega}}]{Morbidelli2016}
{Morbidelli}, A., {Bitsch}, B., {Crida}, A., {et~al.} 2016,
  \href{http://dx.doi.org/10.1016/j.icarus.2015.11.027}{\color{magenta}\icarus},
  \href{https://ui.adsabs.harvard.edu/abs/2016Icar..267..368M}{\color{blue}267},
  \href{https://ui.adsabs.harvard.edu/abs/2016Icar..267..368M}{\color{blue}368}

\bibitem[{{M{\"u}ller} {et~al.}(2020){M{\"u}ller}, {Ben-Yami}, \&
  {Helled}}]{ller2020ApJ}
{M{\"u}ller}, S., {Ben-Yami}, M., \& {Helled}, R. 2020,
  \href{http://dx.doi.org/10.3847/1538-4357/abba19}{\color{magenta}\apj},
  \href{https://ui.adsabs.harvard.edu/abs/2020ApJ...903..147M}{\color{blue}903},
  \href{https://ui.adsabs.harvard.edu/abs/2020ApJ...903..147M}{\color{blue}147}

\bibitem[{{M{\"u}ller} \& {Helled}(2023)}]{ller2023FrASS}
{M{\"u}ller}, S., \& {Helled}, R. 2023,
  \href{http://dx.doi.org/10.3389/fspas.2023.1179000}{\color{magenta}FrASS},
  \href{https://ui.adsabs.harvard.edu/abs/2023FrASS..1079000M}{\color{blue}10},
  \href{https://ui.adsabs.harvard.edu/abs/2023FrASS..1079000M}{\color{blue}1179000}

\bibitem[{{Muro-Arena} {et~al.}(2020){Muro-Arena}, {Benisty}, {Ginski},
  {Dominik}, {Facchini}, {Villenave}, {van Boekel}, {Chauvin}, {Garufi},
  {Henning}, {Janson}, {Keppler}, {Matter}, {M{\'e}nard}, {Stolker}, {Zurlo},
  {Blanchard}, {Maurel}, {Moeller-Nilsson}, {Petit}, {Roux}, {Sevin}, \&
  {Wildi}}]{MuroArena2020}
{Muro-Arena}, G.~A., {Benisty}, M., {Ginski}, C., {et~al.} 2020,
  \href{http://dx.doi.org/10.1051/0004-6361/201936509}{\color{magenta}\aap},
  \href{https://ui.adsabs.harvard.edu/abs/2020A&A...635A.121M}{\color{blue}635},
  \href{https://ui.adsabs.harvard.edu/abs/2020A&A...635A.121M}{\color{blue}A121}

\bibitem[{{Olofsson} {et~al.}(2022){Olofsson}, {Th{\'e}bault}, {Kennedy}, \&
  {Bayo}}]{Olofsson2022A&A}
{Olofsson}, J., {Th{\'e}bault}, P., {Kennedy}, G.~M., \& {Bayo}, A. 2022,
  \href{http://dx.doi.org/10.1051/0004-6361/202243794}{\color{magenta}\aap},
  \href{https://ui.adsabs.harvard.edu/abs/2022A&A...664A.122O}{\color{blue}664},
  \href{https://ui.adsabs.harvard.edu/abs/2022A&A...664A.122O}{\color{blue}A122}

\bibitem[{{Olofsson} {et~al.}(2018){Olofsson}, {van Holstein}, {Boccaletti},
  {Janson}, {Th{\'e}bault}, {Gratton}, {Lazzoni}, {Kral}, {Bayo}, {Canovas},
  {Caceres}, {Ginski}, {Pinte}, {Asensio-Torres}, {Chauvin}, {Desidera},
  {Henning}, {Langlois}, {Milli}, {Schlieder}, {Schreiber}, {Augereau},
  {Bonnefoy}, {Buenzli}, {Brandner}, {Durkan}, {Engler}, {Feldt}, {Godoy},
  {Grady}, {Hagelberg}, {Lagrange}, {Lannier}, {Ligi}, {Maire}, {Mawet},
  {M{\'e}nard}, {Mesa}, {Mouillet}, {Peretti}, {Perrot}, {Salter}, {Schmidt},
  {Sissa}, {Thalmann}, {Vigan}, {Abe}, {Feautrier}, {Le Mignant}, {Moulin},
  {Pavlov}, {Rabou}, {Rousset}, \& {Roux}}]{Olofsson2018}
{Olofsson}, J., {van Holstein}, R.~G., {Boccaletti}, A., {et~al.} 2018,
  \href{http://dx.doi.org/10.1051/0004-6361/201832583}{\color{magenta}\aap},
  \href{https://ui.adsabs.harvard.edu/abs/2018A&A...617A.109O}{\color{blue}617},
  \href{https://ui.adsabs.harvard.edu/abs/2018A&A...617A.109O}{\color{blue}A109}

\bibitem[{{Olofsson} {et~al.}(2023){Olofsson}, {Th{\'e}bault}, {Bayo}, {Milli},
  {van Holstein}, {Henning}, {Medina-Olea}, {Godoy}, \&
  {Mauc{\'o}}}]{Olofsson2023A&A}
{Olofsson}, J., {Th{\'e}bault}, P., {Bayo}, A., {et~al.} 2023,
  \href{http://dx.doi.org/10.1051/0004-6361/202346097}{\color{magenta}\aap},
  \href{https://ui.adsabs.harvard.edu/abs/2023A&A...674A..84O}{\color{blue}674},
  \href{https://ui.adsabs.harvard.edu/abs/2023A&A...674A..84O}{\color{blue}A84}

\bibitem[{{Pacetti} {et~al.}(2022){Pacetti}, {Turrini}, {Schisano}, {Molinari},
  {Fonte}, {Politi}, {Hennebelle}, {Klessen}, {Testi}, \&
  {Lebreuilly}}]{Pacetti2022ApJ}
{Pacetti}, E., {Turrini}, D., {Schisano}, E., {et~al.} 2022,
  \href{http://dx.doi.org/10.3847/1538-4357/ac8b11}{\color{magenta}\apj},
  \href{https://ui.adsabs.harvard.edu/abs/2022ApJ...937...36P}{\color{blue}937},
  \href{https://ui.adsabs.harvard.edu/abs/2022ApJ...937...36P}{\color{blue}36}

\bibitem[{{Pecaut} {et~al.}(2012){Pecaut}, {Mamajek}, \& {Bubar}}]{Pecaut2012}
{Pecaut}, M.~J., {Mamajek}, E.~E., \& {Bubar}, E.~J. 2012,
  \href{http://dx.doi.org/10.1088/0004-637X/746/2/154}{\color{magenta}\apj},
  \href{https://ui.adsabs.harvard.edu/abs/2012ApJ...746..154P}{\color{blue}746},
  \href{https://ui.adsabs.harvard.edu/abs/2012ApJ...746..154P}{\color{blue}154}

\bibitem[{{Pinilla} {et~al.}(2012){Pinilla}, {Benisty}, \&
  {Birnstiel}}]{Pinilla2012A&A}
{Pinilla}, P., {Benisty}, M., \& {Birnstiel}, T. 2012,
  \href{http://dx.doi.org/10.1051/0004-6361/201219315}{\color{magenta}\aap},
  \href{https://ui.adsabs.harvard.edu/abs/2012A&A...545A..81P}{\color{blue}545},
  \href{https://ui.adsabs.harvard.edu/abs/2012A&A...545A..81P}{\color{blue}A81}

\bibitem[{{Pinilla} {et~al.}(2018{\natexlab{a}}){Pinilla}, {Tazzari},
  {Pascucci}, {Youdin}, {Garufi}, {Manara}, {Testi}, {van der Plas},
  {Barenfeld}, {Canovas}, {Cox}, {Hendler}, {P{\'e}rez}, \& {van der
  Marel}}]{Pinilla_2018}
{Pinilla}, P., {Tazzari}, M., {Pascucci}, I., {et~al.} 2018{\natexlab{a}},
  \href{http://dx.doi.org/10.3847/1538-4357/aabf94}{\color{magenta}\apj},
  \href{https://ui.adsabs.harvard.edu/abs/2018ApJ...859...32P}{\color{blue}859},
  \href{https://ui.adsabs.harvard.edu/abs/2018ApJ...859...32P}{\color{blue}32}

\bibitem[{{Pinilla} {et~al.}(2018{\natexlab{b}}){Pinilla}, {Benisty}, {de
  Boer}, {Manara}, {Bouvier}, {Dominik}, {Ginski}, {Loomis}, \& {Sicilia
  Aguilar}}]{Pinilla2018}
{Pinilla}, P., {Benisty}, M., {de Boer}, J., {et~al.} 2018{\natexlab{b}},
  \href{http://dx.doi.org/10.3847/1538-4357/aae824}{\color{magenta}\apj},
  \href{https://ui.adsabs.harvard.edu/abs/2018ApJ...868...85P}{\color{blue}868},
  \href{https://ui.adsabs.harvard.edu/abs/2018ApJ...868...85P}{\color{blue}85}

\bibitem[{{Piso} \& {Youdin}(2014)}]{Piso2014}
{Piso}, A.-M.~A., \& {Youdin}, A.~N. 2014,
  \href{http://dx.doi.org/10.1088/0004-637X/786/1/21}{\color{magenta}\apj},
  \href{https://ui.adsabs.harvard.edu/abs/2014ApJ...786...21P}{\color{blue}786},
  \href{https://ui.adsabs.harvard.edu/abs/2014ApJ...786...21P}{\color{blue}21}

\bibitem[{{Piso} {et~al.}(2015){Piso}, {Youdin}, \& {Murray-Clay}}]{Piso2015}
{Piso}, A.-M.~A., {Youdin}, A.~N., \& {Murray-Clay}, R.~A. 2015,
  \href{http://dx.doi.org/10.1088/0004-637X/800/2/82}{\color{magenta}\apj},
  \href{https://ui.adsabs.harvard.edu/abs/2015ApJ...800...82P}{\color{blue}800},
  \href{https://ui.adsabs.harvard.edu/abs/2015ApJ...800...82P}{\color{blue}82}

\bibitem[{{Plotnykov} \& {Valencia}(2020)}]{Plotnykov2020MNRAS}
{Plotnykov}, M., \& {Valencia}, D. 2020,
  \href{http://dx.doi.org/10.1093/mnras/staa2615}{\color{magenta}\mnras},
  \href{https://ui.adsabs.harvard.edu/abs/2020MNRAS.499..932P}{\color{blue}499},
  \href{https://ui.adsabs.harvard.edu/abs/2020MNRAS.499..932P}{\color{blue}932}

\bibitem[{{Poch} {et~al.}(2023){Poch}, {Istiqomah}, {Quirico}, {Beck},
  {Schmitt}, {Theul{\'e}}, {Faure}, {Hily-Blant}, {Bonal}, {Raponi},
  {Ciarniello}, {Rousseau}, {Potin}, {Brissaud}, {Flandinet}, {Filacchione},
  {Pommerol}, {Thomas}, {Kappel}, {Mennella}, {Moroz}, {Vinogradoff}, {Arnold},
  {Erard}, {Bockel{\'e}e-Morvan}, {Leyrat}, {Capaccioni}, {De Sanctis},
  {Longobardo}, {Mancarella}, {Palomba}, \& {Tosi}}]{Poch23}
{Poch}, O., {Istiqomah}, I., {Quirico}, E., {et~al.} 2023,
  \href{http://dx.doi.org/10.1007/978-3-031-29003-9_31}{\color{magenta}European
  Conference on Laboratory Astrophysics ECLA2020. The Interplay of Dust},
  \href{https://ui.adsabs.harvard.edu/abs/2023ASSP...59..271P}{\color{blue}59},
  \href{https://ui.adsabs.harvard.edu/abs/2023ASSP...59..271P}{\color{blue}271}

\bibitem[{{Pollack} {et~al.}(1996){Pollack}, {Hubickyj}, {Bodenheimer},
  {Lissauer}, {Podolak}, \& {Greenzweig}}]{Pollack1996}
{Pollack}, J.~B., {Hubickyj}, O., {Bodenheimer}, P., {et~al.} 1996,
  \href{http://dx.doi.org/10.1006/icar.1996.0190}{\color{magenta}\icarus},
  \href{https://ui.adsabs.harvard.edu/abs/1996Icar..124...62P}{\color{blue}124},
  \href{https://ui.adsabs.harvard.edu/abs/1996Icar..124...62P}{\color{blue}62}

\bibitem[{Popa \& Udrea(2019)}]{s19092076}
Popa, D., \& Udrea, F. 2019,
  \href{http://dx.doi.org/10.3390/s19092076}{\color{magenta}Sensors}, 19

\bibitem[{{Preibisch} \& {Feigelson}(2005)}]{Preibisch2005}
{Preibisch}, T., \& {Feigelson}, E.~D. 2005,
  \href{http://dx.doi.org/10.1086/432094}{\color{magenta}\apjs},
  \href{https://ui.adsabs.harvard.edu/abs/2005ApJS..160..390P}{\color{blue}160},
  \href{https://ui.adsabs.harvard.edu/abs/2005ApJS..160..390P}{\color{blue}390}

\bibitem[{{Quiroz} {et~al.}(2022){Quiroz}, {Wallack}, {Ren}, {Dong}, {Xuan},
  {Mawet}, {Millar-Blanchaer}, \& {Ruane}}]{Quiroz2022ApJ}
{Quiroz}, J., {Wallack}, N.~L., {Ren}, B., {et~al.} 2022,
  \href{http://dx.doi.org/10.3847/2041-8213/ac3e62}{\color{magenta}\apjl},
  \href{https://ui.adsabs.harvard.edu/abs/2022ApJ...924L...4Q}{\color{blue}924},
  \href{https://ui.adsabs.harvard.edu/abs/2022ApJ...924L...4Q}{\color{blue}L4}

\bibitem[{{Ren} {et~al.}(2019){Ren}, {Choquet}, {Perrin}, {Duch{\^e}ne},
  {Debes}, {Pueyo}, {Rice}, {Chen}, {Schneider}, {Esposito}, {Poteet}, {Wang},
  {Ammons}, {Ansdell}, {Arriaga}, {Bailey}, {Barman}, {Sebasti{\'a}n Bruzzone},
  {Bulger}, {Chilcote}, {Cotten}, {De Rosa}, {Doyon}, {Fitzgerald}, {Follette},
  {Goodsell}, {Gerard}, {Graham}, {Greenbaum}, {Hagan}, {Hibon}, {Hines},
  {Hung}, {Ingraham}, {Kalas}, {Konopacky}, {Larkin}, {Macintosh}, {Maire},
  {Marchis}, {Marois}, {Mazoyer}, {M{\'e}nard}, {Metchev}, {Millar-Blanchaer},
  {Mittal}, {Moerchen}, {Nielsen}, {N'Diaye}, {Oppenheimer}, {Palmer},
  {Patience}, {Pinte}, {Poyneer}, {Rajan}, {Rameau}, {Rantakyr{\"o}}, {Ruffio},
  {Ryan}, {Savransky}, {Schneider}, {Sivaramakrishnan}, {Song}, {Soummer},
  {Stark}, {Thomas}, {Vigan}, {Wallace}, {Ward-Duong}, {Wiktorowicz}, {Wolff},
  {Ygouf}, \& {Norman}}]{ren19}
{Ren}, B., {Choquet}, {\'E}., {Perrin}, M.~D., {et~al.} 2019,
  \href{http://dx.doi.org/10.3847/1538-4357/ab3403}{\color{magenta}\apj},
  \href{https://ui.adsabs.harvard.edu/abs/2019ApJ...882...64R}{\color{blue}882},
  \href{https://ui.adsabs.harvard.edu/abs/2019ApJ...882...64R}{\color{blue}64}

\bibitem[{{Ren}(2023)}]{Ren2023A&A}
{Ren}, B.~B. 2023,
  \href{http://dx.doi.org/10.1051/0004-6361/202347354}{\color{magenta}\aap},
  \href{https://ui.adsabs.harvard.edu/abs/2023A&A...679A..18R}{\color{blue}679},
  \href{https://ui.adsabs.harvard.edu/abs/2023A&A...679A..18R}{\color{blue}A18}

\bibitem[{{Ren} {et~al.}(2023){Ren}, {Rebollido}, {Choquet}, {Zhou}, {Perrin},
  {Schneider}, {Milli}, {Wolff}, {Chen}, {Debes}, {Hagan}, {Hines},
  {Millar-Blanchaer}, {Pueyo}, {Roberge}, {Serabyn}, \&
  {Soummer}}]{Ren23debris}
{Ren}, B.~B., {Rebollido}, I., {Choquet}, {\'E}., {et~al.} 2023,
  \href{http://dx.doi.org/10.1051/0004-6361/202245458}{\color{magenta}\aap},
  \href{https://ui.adsabs.harvard.edu/abs/2023A&A...672A.114R}{\color{blue}672},
  \href{https://ui.adsabs.harvard.edu/abs/2023A&A...672A.114R}{\color{blue}A114}

\bibitem[{{Rice} {et~al.}(2006){Rice}, {Armitage}, {Wood}, \&
  {Lodato}}]{Rice2006}
{Rice}, W.~K.~M., {Armitage}, P.~J., {Wood}, K., \& {Lodato}, G. 2006,
  \href{http://dx.doi.org/10.1111/j.1365-2966.2006.11113.x}{\color{magenta}\mnras},
  \href{https://ui.adsabs.harvard.edu/abs/2006MNRAS.373.1619R}{\color{blue}373},
  \href{https://ui.adsabs.harvard.edu/abs/2006MNRAS.373.1619R}{\color{blue}1619}

\bibitem[{{Rodigas} {et~al.}(2014){Rodigas}, {Malhotra}, \&
  {Hinz}}]{Rodigas2014ApJ}
{Rodigas}, T.~J., {Malhotra}, R., \& {Hinz}, P.~M. 2014,
  \href{http://dx.doi.org/10.1088/0004-637X/780/1/65}{\color{magenta}\apj},
  \href{https://ui.adsabs.harvard.edu/abs/2014ApJ...780...65R}{\color{blue}780},
  \href{https://ui.adsabs.harvard.edu/abs/2014ApJ...780...65R}{\color{blue}65}

\bibitem[{{Rodigas} {et~al.}(2015){Rodigas}, {Stark}, {Weinberger}, {Debes},
  {Hinz}, {Close}, {Chen}, {Smith}, {Males}, {Skemer}, {Puglisi}, {Follette},
  {Morzinski}, {Wu}, {Briguglio}, {Esposito}, {Pinna}, {Riccardi}, {Schneider},
  \& {Xompero}}]{Rodigas2015}
{Rodigas}, T.~J., {Stark}, C.~C., {Weinberger}, A., {et~al.} 2015,
  \href{http://dx.doi.org/10.1088/0004-637X/798/2/96}{\color{magenta}\apj},
  \href{https://ui.adsabs.harvard.edu/abs/2015ApJ...798...96R}{\color{blue}798},
  \href{https://ui.adsabs.harvard.edu/abs/2015ApJ...798...96R}{\color{blue}96}

\bibitem[{{Rogers} \& {Seager}(2010)}]{Rogers2010ApJ}
{Rogers}, L.~A., \& {Seager}, S. 2010,
  \href{http://dx.doi.org/10.1088/0004-637X/716/2/1208}{\color{magenta}\apj},
  \href{https://ui.adsabs.harvard.edu/abs/2010ApJ...716.1208R}{\color{blue}716},
  \href{https://ui.adsabs.harvard.edu/abs/2010ApJ...716.1208R}{\color{blue}1208}

\bibitem[{{Ruane} {et~al.}(2019){Ruane}, {Ngo}, {Mawet}, {Absil}, {Choquet},
  {Cook}, {Gomez Gonzalez}, {Huby}, {Matthews}, {Meshkat}, {Reggiani},
  {Serabyn}, {Wallack}, \& {Xuan}}]{Ruane2019}
{Ruane}, G., {Ngo}, H., {Mawet}, D., {et~al.} 2019,
  \href{http://dx.doi.org/10.3847/1538-3881/aafee2}{\color{magenta}\aj},
  \href{https://ui.adsabs.harvard.edu/abs/2019AJ....157..118R}{\color{blue}157},
  \href{https://ui.adsabs.harvard.edu/abs/2019AJ....157..118R}{\color{blue}118}

\bibitem[{Rupprecht(2005)}]{rupprecht2005requirements}
Rupprecht, G. 2005, Requirements for Scientific Instruments on the VLT Unit
  Telescopes, Tech. rep., VLT-SPE-ESO-10000-2723 publicly available at
  \url{https://www.eso.org/sci/facilities/develop/documents/VLT-SPE-ESO-10000-2723_is1.pdf}

\bibitem[{{Service} {et~al.}(2016){Service}, {Lu}, {Campbell}, {Sitarski},
  {Ghez}, \& {Anderson}}]{Service2016}
{Service}, M., {Lu}, J.~R., {Campbell}, R., {et~al.} 2016,
  \href{http://dx.doi.org/10.1088/1538-3873/128/967/095004}{\color{magenta}\pasp},
  \href{https://ui.adsabs.harvard.edu/abs/2016PASP..128i5004S}{\color{blue}128},
  \href{https://ui.adsabs.harvard.edu/abs/2016PASP..128i5004S}{\color{blue}095004}

\bibitem[{{Sicilia-Aguilar} {et~al.}(2020){Sicilia-Aguilar}, {Manara}, {de
  Boer}, {Benisty}, {Pinilla}, \& {Bouvier}}]{Sicilia2020}
{Sicilia-Aguilar}, A., {Manara}, C.~F., {de Boer}, J., {et~al.} 2020,
  \href{http://dx.doi.org/10.1051/0004-6361/201936565}{\color{magenta}\aap},
  \href{https://ui.adsabs.harvard.edu/abs/2020A&A...633A..37S}{\color{blue}633},
  \href{https://ui.adsabs.harvard.edu/abs/2020A&A...633A..37S}{\color{blue}A37}

\bibitem[{{Singh} {et~al.}(2021){Singh}, {Bhowmik}, {Boccaletti},
  {Th{\'e}bault}, {Kral}, {Milli}, {Mazoyer}, {Pantin}, {van Holstein},
  {Olofsson}, {Boukrouche}, {Di Folco}, {Janson}, {Langlois}, {Maire}, {Vigan},
  {Benisty}, {Augereau}, {Perrot}, {Gratton}, {Henning}, {M{\'e}nard},
  {Rickman}, {Wahhaj}, {Zurlo}, {Biller}, {Bonnefoy}, {Chauvin}, {Delorme},
  {Desidera}, {D'Orazi}, {Feldt}, {Hagelberg}, {Keppler}, {Kopytova},
  {Lagadec}, {Lagrange}, {Mesa}, {Meyer}, {Rouan}, {Sissa}, {Schmidt},
  {Jaquet}, {Fusco}, {Pavlov}, \& {Rabou}}]{Singh2021A&A}
{Singh}, G., {Bhowmik}, T., {Boccaletti}, A., {et~al.} 2021,
  \href{http://dx.doi.org/10.1051/0004-6361/202140319}{\color{magenta}\aap},
  \href{https://ui.adsabs.harvard.edu/abs/2021A&A...653A..79S}{\color{blue}653},
  \href{https://ui.adsabs.harvard.edu/abs/2021A&A...653A..79S}{\color{blue}A79}

\bibitem[{{Sitko} {et~al.}(2012){Sitko}, {Day}, {Kimes}, {Beerman}, {Martus},
  {Lynch}, {Russell}, {Grady}, {Schneider}, {Lisse}, {Nuth}, {Cur{\'e}},
  {Henden}, {Kraus}, {Motta}, {Tamura}, {Hornbeck}, {Williger}, \&
  {Fugazza}}]{Sitko2012ApJ}
{Sitko}, M.~L., {Day}, A.~N., {Kimes}, R.~L., {et~al.} 2012,
  \href{http://dx.doi.org/10.1088/0004-637X/745/1/29}{\color{magenta}\apj},
  \href{https://ui.adsabs.harvard.edu/abs/2012ApJ...745...29S}{\color{blue}745},
  \href{https://ui.adsabs.harvard.edu/abs/2012ApJ...745...29S}{\color{blue}29}

\bibitem[{{Soummer} {et~al.}(2012){Soummer}, {Pueyo}, \&
  {Larkin}}]{Soummer2012}
{Soummer}, R., {Pueyo}, L., \& {Larkin}, J. 2012,
  \href{http://dx.doi.org/10.1088/2041-8205/755/2/L28}{\color{magenta}\apjl},
  \href{https://ui.adsabs.harvard.edu/abs/2012ApJ...755L..28S}{\color{blue}755},
  \href{https://ui.adsabs.harvard.edu/abs/2012ApJ...755L..28S}{\color{blue}L28}

\bibitem[{{Spiegel} \& {Burrows}(2012)}]{Spiegel2012ApJ}
{Spiegel}, D.~S., \& {Burrows}, A. 2012,
  \href{http://dx.doi.org/10.1088/0004-637X/745/2/174}{\color{magenta}\apj},
  \href{https://ui.adsabs.harvard.edu/abs/2012ApJ...745..174S}{\color{blue}745},
  \href{https://ui.adsabs.harvard.edu/abs/2012ApJ...745..174S}{\color{blue}174}

\bibitem[{{Stadler} {et~al.}(2023){Stadler}, {Benisty}, {Izquierdo},
  {Facchini}, {Teague}, {Kurtovic}, {Pinilla}, {Bae}, {Ansdell}, {Loomis},
  {Mayama}, {Perez}, \& {Testi}}]{Stadler2023}
{Stadler}, J., {Benisty}, M., {Izquierdo}, A., {et~al.} 2023,
  \href{http://dx.doi.org/10.1051/0004-6361/202245381}{\color{magenta}\aap},
  \href{https://ui.adsabs.harvard.edu/abs/2023A&A...670L...1S}{\color{blue}670},
  \href{https://ui.adsabs.harvard.edu/abs/2023A&A...670L...1S}{\color{blue}L1}

\bibitem[{{Stark} {et~al.}(2023){Stark}, {Ren}, {MacGregor}, {Howard}, {Hurt},
  {Weinberger}, {Schneider}, \& {Choquet}}]{Stark2023ApJ}
{Stark}, C.~C., {Ren}, B., {MacGregor}, M.~A., {et~al.} 2023,
  \href{http://dx.doi.org/10.3847/1538-4357/acbb64}{\color{magenta}\apj},
  \href{https://ui.adsabs.harvard.edu/abs/2023ApJ...945..131S}{\color{blue}945},
  \href{https://ui.adsabs.harvard.edu/abs/2023ApJ...945..131S}{\color{blue}131}

\bibitem[{{Stasevic} {et~al.}(2023){Stasevic}, {Milli}, {Mazoyer}, {Lagrange},
  {Bonnefoy}, {Faramaz-Gorka}, {M{\'e}nard}, {Boccaletti}, {Choquet}, {Shuai},
  {Olofsson}, {Chomez}, {Ren}, {Rubini}, {Desgrange}, {Gratton}, {Chauvin},
  {Vigan}, \& {Matthews}}]{Stasevic2023A&A}
{Stasevic}, S., {Milli}, J., {Mazoyer}, J., {et~al.} 2023,
  \href{http://dx.doi.org/10.1051/0004-6361/202346720}{\color{magenta}\aap},
  \href{https://ui.adsabs.harvard.edu/abs/2023A&A...678A...8S}{\color{blue}678},
  \href{https://ui.adsabs.harvard.edu/abs/2023A&A...678A...8S}{\color{blue}A8}

\bibitem[{{Stolker} {et~al.}(2016){Stolker}, {Dominik}, {Avenhaus}, {Min}, {de
  Boer}, {Ginski}, {Schmid}, {Juhasz}, {Bazzon}, {Waters}, {Garufi},
  {Augereau}, {Benisty}, {Boccaletti}, {Henning}, {Langlois}, {Maire},
  {M{\'e}nard}, {Meyer}, {Pinte}, {Quanz}, {Thalmann}, {Beuzit}, {Carbillet},
  {Costille}, {Dohlen}, {Feldt}, {Gisler}, {Mouillet}, {Pavlov}, {Perret},
  {Petit}, {Pragt}, {Rochat}, {Roelfsema}, {Salasnich}, {Soenke}, \&
  {Wildi}}]{Stolker2016A&A}
{Stolker}, T., {Dominik}, C., {Avenhaus}, H., {et~al.} 2016,
  \href{http://dx.doi.org/10.1051/0004-6361/201528039}{\color{magenta}\aap},
  \href{https://ui.adsabs.harvard.edu/abs/2016A&A...595A.113S}{\color{blue}595},
  \href{https://ui.adsabs.harvard.edu/abs/2016A&A...595A.113S}{\color{blue}A113}

\bibitem[{{Takeuchi} \& {Artymowicz}(2001)}]{Takeuchi2001ApJ}
{Takeuchi}, T., \& {Artymowicz}, P. 2001,
  \href{http://dx.doi.org/10.1086/322252}{\color{magenta}\apj},
  \href{https://ui.adsabs.harvard.edu/abs/2001ApJ...557..990T}{\color{blue}557},
  \href{https://ui.adsabs.harvard.edu/abs/2001ApJ...557..990T}{\color{blue}990}

\bibitem[{{Th{\'e}bault}(2009)}]{bault2009}
{Th{\'e}bault}, P. 2009,
  \href{http://dx.doi.org/10.1051/0004-6361/200912396}{\color{magenta}\aap},
  \href{https://ui.adsabs.harvard.edu/abs/2009A&A...505.1269T}{\color{blue}505},
  \href{https://ui.adsabs.harvard.edu/abs/2009A&A...505.1269T}{\color{blue}1269}

\bibitem[{{Thorngren} {et~al.}(2016){Thorngren}, {Fortney}, {Murray-Clay}, \&
  {Lopez}}]{Thorngren2016}
{Thorngren}, D.~P., {Fortney}, J.~J., {Murray-Clay}, R.~A., \& {Lopez}, E.~D.
  2016,
  \href{http://dx.doi.org/10.3847/0004-637X/831/1/64}{\color{magenta}\apj},
  \href{https://ui.adsabs.harvard.edu/abs/2016ApJ...831...64T}{\color{blue}831},
  \href{https://ui.adsabs.harvard.edu/abs/2016ApJ...831...64T}{\color{blue}64}

\bibitem[{{Turrini} {et~al.}(2021){Turrini}, {Schisano}, {Fonte}, {Molinari},
  {Politi}, {Fedele}, {Pani{\'c}}, {Kama}, {Changeat}, \&
  {Tinetti}}]{Turrini2021ApJ}
{Turrini}, D., {Schisano}, E., {Fonte}, S., {et~al.} 2021,
  \href{http://dx.doi.org/10.3847/1538-4357/abd6e5}{\color{magenta}\apj},
  \href{https://ui.adsabs.harvard.edu/abs/2021ApJ...909...40T}{\color{blue}909},
  \href{https://ui.adsabs.harvard.edu/abs/2021ApJ...909...40T}{\color{blue}40}

\bibitem[{{van der Marel} {et~al.}(2015){van der Marel}, {van Dishoeck},
  {Bruderer}, {P{\'e}rez}, \& {Isella}}]{Marel2015}
{van der Marel}, N., {van Dishoeck}, E.~F., {Bruderer}, S., {et~al.} 2015,
  \href{http://dx.doi.org/10.1051/0004-6361/201525658}{\color{magenta}\aap},
  \href{https://ui.adsabs.harvard.edu/abs/2015A&A...579A.106V}{\color{blue}579},
  \href{https://ui.adsabs.harvard.edu/abs/2015A&A...579A.106V}{\color{blue}A106}

\bibitem[{{Vigan}(2020)}]{Vigan2020ascl}
{Vigan}, A. 2020, {vlt-sphere: Automatic VLT/SPHERE data reduction and
  analysis}, Astrophysics Source Code Library, record ascl:2009.002,
  Astrophysics Source Code Library, record ascl:2009.002

\bibitem[{{Vigan} {et~al.}(2010){Vigan}, {Moutou}, {Langlois}, {Allard},
  {Boccaletti}, {Carbillet}, {Mouillet}, \& {Smith}}]{Vigan2010}
{Vigan}, A., {Moutou}, C., {Langlois}, M., {et~al.} 2010,
  \href{http://dx.doi.org/10.1111/j.1365-2966.2010.16916.x}{\color{magenta}\mnras},
  \href{https://ui.adsabs.harvard.edu/abs/2010MNRAS.407...71V}{\color{blue}407},
  \href{https://ui.adsabs.harvard.edu/abs/2010MNRAS.407...71V}{\color{blue}71}

\bibitem[{Wahhaj {et~al.}(2022)Wahhaj, Jones, de~Rosa, Rodler, van~den Ancker,
  \& Boffin}]{SPHEREManual}
Wahhaj, Z., Jones, M., de~Rosa, R., {et~al.} 2022, Very Large Telescope SPHERE
  User Manual, Tech. rep., VLT-MAN-SPH-14690-0430 publicaly available at
  \url{https://www.eso.org/sci/facilities/paranal/instruments/sphere/doc/VLT-MAN-SPH-14690-0430_P111_dec_2022_zwa.pdf}

\bibitem[{{Wang} {et~al.}(2020{\natexlab{a}}){Wang}, {Wang}, {Ma}, {Chilcote},
  {Ertel}, {Guyon}, {Ilyin}, {Jovanovic}, {Kalas}, {Lozi}, {Macintosh},
  {Strassmeier}, \& {Stone}}]{2020wangji}
{Wang}, J., {Wang}, J.~J., {Ma}, B., {et~al.} 2020{\natexlab{a}},
  \href{http://dx.doi.org/10.3847/1538-3881/ababa7}{\color{magenta}\aj},
  \href{https://ui.adsabs.harvard.edu/abs/2020AJ....160..150W}{\color{blue}160},
  \href{https://ui.adsabs.harvard.edu/abs/2020AJ....160..150W}{\color{blue}150}

\bibitem[{{Wang} {et~al.}(2020{\natexlab{b}}){Wang}, {Ginzburg}, {Ren},
  {Wallack}, {Gao}, {Mawet}, {Bond}, {Cetre}, {Wizinowich}, {De Rosa}, {Ruane},
  {Liu}, {Absil}, {Alvarez}, {Baranec}, {Choquet}, {Chun}, {Defr{\`e}re},
  {Delorme}, {Duch{\^e}ne}, {Forsberg}, {Ghez}, {Guyon}, {Hall}, {Huby},
  {Jolivet}, {Jensen-Clem}, {Jovanovic}, {Karlsson}, {Lilley}, {Matthews},
  {M{\'e}nard}, {Meshkat}, {Millar-Blanchaer}, {Ngo}, {Orban de Xivry},
  {Pinte}, {Ragland}, {Serabyn}, {Catal{\'a}n}, {Wang}, {Wetherell},
  {Williams}, {Ygouf}, \& {Zuckerman}}]{Wang2020AJ}
{Wang}, J.~J., {Ginzburg}, S., {Ren}, B., {et~al.} 2020{\natexlab{b}},
  \href{http://dx.doi.org/10.3847/1538-3881/ab8aef}{\color{magenta}\aj},
  \href{https://ui.adsabs.harvard.edu/abs/2020AJ....159..263W}{\color{blue}159},
  \href{https://ui.adsabs.harvard.edu/abs/2020AJ....159..263W}{\color{blue}263}

\bibitem[{{Weiss} \& {Marcy}(2014)}]{Weiss2014ApJ}
{Weiss}, L.~M., \& {Marcy}, G.~W. 2014,
  \href{http://dx.doi.org/10.1088/2041-8205/783/1/L6}{\color{magenta}\apjl},
  \href{https://ui.adsabs.harvard.edu/abs/2014ApJ...783L...6W}{\color{blue}783},
  \href{https://ui.adsabs.harvard.edu/abs/2014ApJ...783L...6W}{\color{blue}L6}

\bibitem[{{Woitke} {et~al.}(2019){Woitke}, {Kamp}, {Antonellini}, {Anthonioz},
  {Baldovin-Saveedra}, {Carmona}, {Dionatos}, {Dominik}, {Greaves},
  {G{\"u}del}, {Ilee}, {Liebhardt}, {Menard}, {Min}, {Pinte}, {Rab}, {Rigon},
  {Thi}, {Thureau}, \& {Waters}}]{Woitke2019PASP}
{Woitke}, P., {Kamp}, I., {Antonellini}, S., {et~al.} 2019,
  \href{http://dx.doi.org/10.1088/1538-3873/aaf4e5}{\color{magenta}\pasp},
  \href{https://ui.adsabs.harvard.edu/abs/2019PASP..131f4301W}{\color{blue}131},
  \href{https://ui.adsabs.harvard.edu/abs/2019PASP..131f4301W}{\color{blue}064301}

\bibitem[{{Xie} {et~al.}(2022){Xie}, {Choquet}, {Vigan}, {Cantalloube},
  {Benisty}, {Boccaletti}, {Bonnefoy}, {Desgrange}, {Garufi}, {Girard},
  {Hagelberg}, {Janson}, {Kenworthy}, {Lagrange}, {Langlois}, {Menard}, \&
  {Zurlo}}]{Xie2022}
{Xie}, C., {Choquet}, E., {Vigan}, A., {et~al.} 2022,
  \href{http://dx.doi.org/10.1051/0004-6361/202243379}{\color{magenta}\aap},
  \href{https://ui.adsabs.harvard.edu/abs/2022A&A...666A..32X}{\color{blue}666},
  \href{https://ui.adsabs.harvard.edu/abs/2022A&A...666A..32X}{\color{blue}A32}

\bibitem[{{Xuan} {et~al.}(2018){Xuan}, {Mawet}, {Ngo}, {Ruane}, {Bailey},
  {Choquet}, {Absil}, {Alvarez}, {Bryan}, {Cook}, {Femen{\'\i}a Castell{\'a}},
  {Gomez Gonzalez}, {Huby}, {Knutson}, {Matthews}, {Ragland}, {Serabyn}, \&
  {Zawol}}]{Xuan2018AJ}
{Xuan}, W.~J., {Mawet}, D., {Ngo}, H., {et~al.} 2018,
  \href{http://dx.doi.org/10.3847/1538-3881/aadae6}{\color{magenta}\aj},
  \href{https://ui.adsabs.harvard.edu/abs/2018AJ....156..156X}{\color{blue}156},
  \href{https://ui.adsabs.harvard.edu/abs/2018AJ....156..156X}{\color{blue}156}

\bibitem[{{Zeng} {et~al.}(2019){Zeng}, {Jacobsen}, {Sasselov}, {Petaev},
  {Vanderburg}, {Lopez-Morales}, {Perez-Mercader}, {Mattsson}, {Li}, {Heising},
  {Bonomo}, {Damasso}, {Berger}, {Cao}, {Levi}, \& {Wordsworth}}]{Zeng2019}
{Zeng}, L., {Jacobsen}, S.~B., {Sasselov}, D.~D., {et~al.} 2019,
  \href{http://dx.doi.org/10.1073/pnas.1812905116}{\color{magenta}PNAS},
  \href{https://ui.adsabs.harvard.edu/abs/2019PNAS..116.9723Z}{\color{blue}116},
  \href{https://ui.adsabs.harvard.edu/abs/2019PNAS..116.9723Z}{\color{blue}9723}

\end{thebibliography}

\begin{appendix}
\section{Auxillary images}\label{sec-maskfig}
\subsection{The mask}

\begin{figure*}[htb!]
  \centering
  \includegraphics[width=1.0\textwidth]{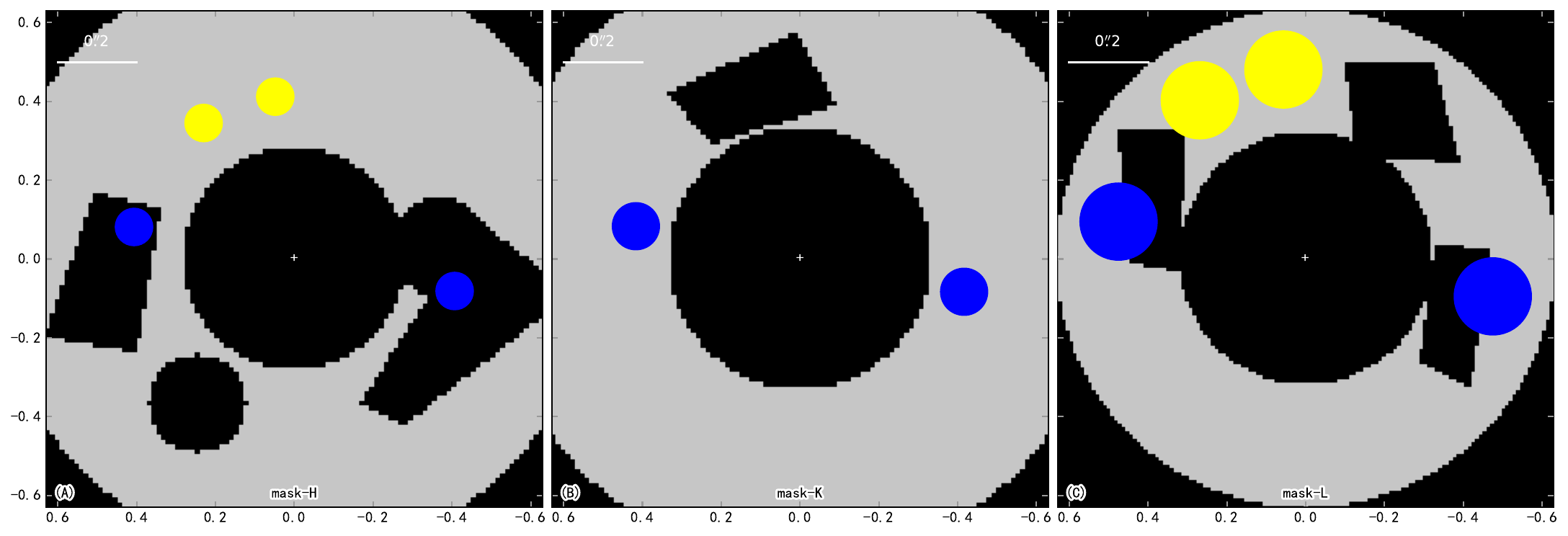}
  \caption{The masks used in forward modeling. The black region was masked excluded when calculating the likelihood function in Eq.~\eqref{eq-disk}. The blue circle masked the region where we calculated average disk surface brightness for disk in the major axis, and the yellow circle masked the region where we calculated average disk surface brightness for disk out of  the major axis to avoid the affect of the shadow.} \label{fig:mask-all}
\end{figure*}

In order to reduce the influence of shadows on disk modeling \citep[e.g.,][]{Pinilla2018}, we selectively excluded specific regions when calculating the likelihood function and calculating the throughput factors. This exclusion was achieved through the application of boolean masks in Fig.~\ref{fig:mask-all}. Similarly, we extracted the disk reflectance at different regions to validate the measurements.

\subsection{Validation of reflectance measurements}
\begin{figure}[htb!]
  \centering
  \includegraphics[width=0.5\textwidth]{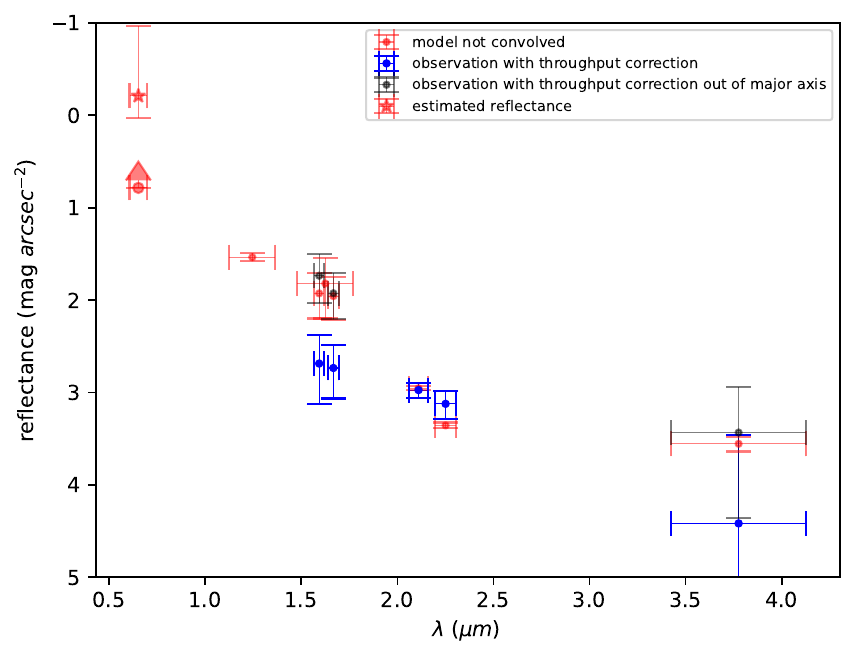}
  \caption{The reflectance of different bands and different methods of J1604.
  Notably, the black and blue points are overlapped with each other in the $K1$ band. 
  We presented the relative reflectance values expressed in units of mag ${\rm arcsec}^{-2}$, where  magnitude is calculated as $-2.5 \log_{10} f_{\rm ref}$ with $f_{\rm ref}$ representing  reflectance ratio. Notably, we obtained the polarized disk intensity instead of the total intensity in the $R$-band, thus we provided  a lower limitation for the reflectance in the $R$-band.
  } \label{fig:J1604reflectance}
\end{figure}

We measured the reflectance using J1604 observation with throughput correction in two distinct regions: the first corresponds to the major axis where the scatting angle is  nearly $90^{\circ}$,% and denoted as $\Gamma$, 
while the second pertains to an off-axis area to reduce the influence of shadows.% and designated as $\Pi$. 
We validated the consistency of the results in Fig.~\ref{fig:J1604reflectance} from different methods and \citet{2023A&AMa}.% obtained from the model by making comparison with the reflectance obtained from observation.

%\subsection{\textcolor{red}{the observation information}}
%\textcolor{red}{ }
%\begin{table*}
% \setlength{\tabcolsep}{1pt}
% \renewcommand{\arraystretch}{1.}
% \caption{Detailed values for disk surface brightness measurement\label{tab:result-aperture}}
% \begin{tabular}{c c r r r r r r r r} \hline\hline

%  &Filters  & $R$ band & $J$ band & $H$ band& $H2$ band & $H3$ band & $K1$ band  & $K2$ band & $L'$ band\\ \hline
% Epochs &2015.06& 2017.08& 2016.07& & & &  & & \\
% Instruments   &VLT/SPHERE& VLT/SPHERE & VLT/SPHERE & VLT/SPHERE & VLT/SPHERE & VLT/SPHERE& VLT/SPHERE & VLT/SPHERE & 95\\ 

% \hline

% \end{tabular}
% \end{table*}
\subsection{MCMC modeling corner plots}\label{sec-corner}

\begin{figure*}[htb!]
  \centering
  \includegraphics[width=1.0\textwidth]{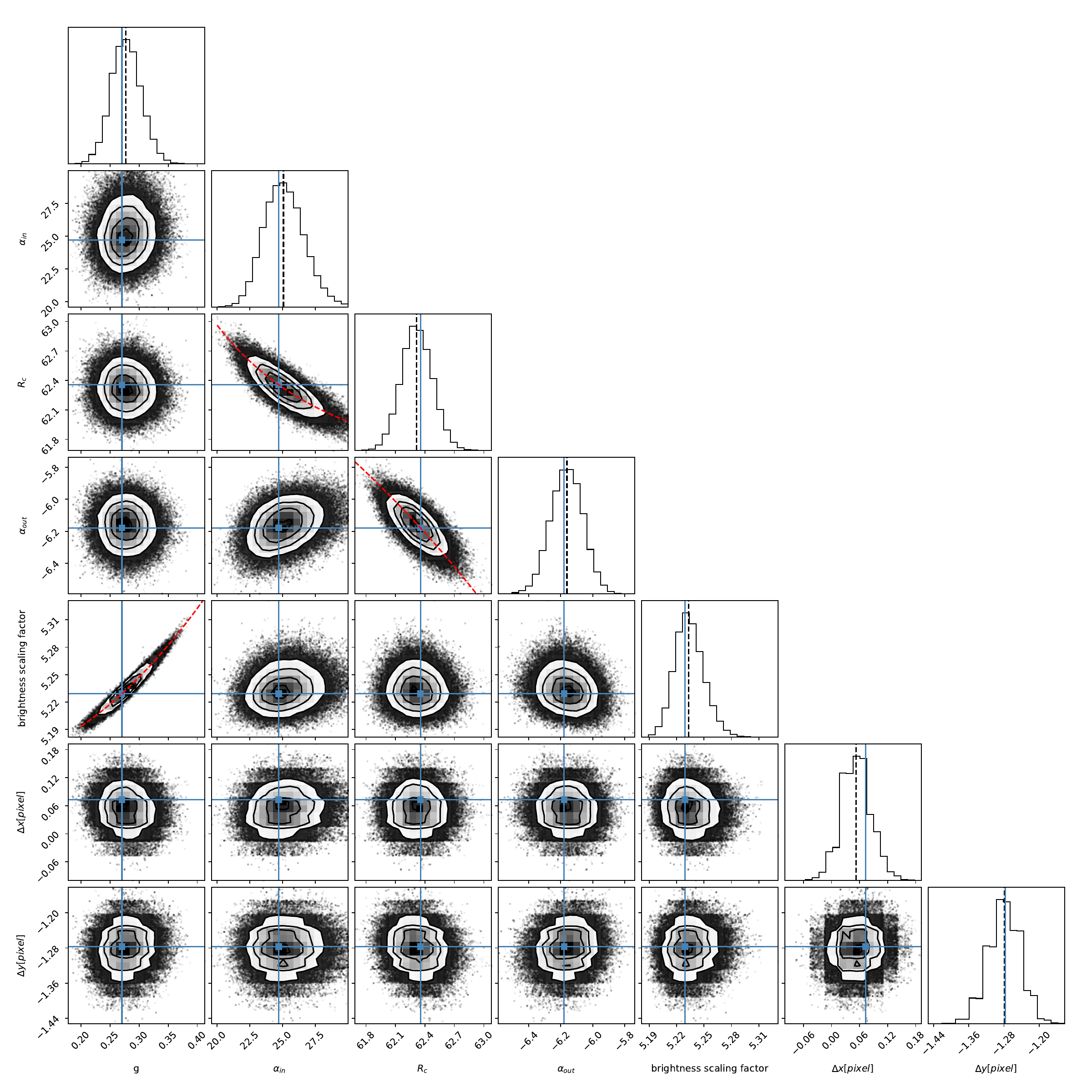}
  \caption{ MCMC posteriors from forward modeling J1604 in $H3$-band.} \label{fig:fig-H3-corner}
\end{figure*}

%\textcolor{red}{
We present the MCMC posteriors for H3-band in Fig.~\ref{fig:fig-H3-corner}. %,~\ref{fig:fig-K1-corner},~\ref{fig:fig-K2-corner},~\ref{fig:fig-L-corner}.
Histograms on the main diagonal show the posterior distributions for each fitted parameter marginalized over all other fitted parameters, with dashed vertical lines showing 50th  percentiles. The maximum likelihood parameters are denoted using blue lines.%}

%\textcolor{red}{
We observed several strong correlations among several parameters (e.g., $g$ and brightness scaling factor, $\alpha_{\rm in}$ and $R_c$, and $\alpha_{\rm out}$ and $R_c$). These correlations are anticipated and can be explained given the scattering phase function and disk geometry setup in Sect.~\ref{sec-forward}. On the one hand, the Henyey--Greenstein parameter $g$ depicts the redistribution of light as a function of scattering angle (i.e., scattering phase function), which distributes more light to smaller scattering angles as a positive $g$ increases in Eq.~\eqref{eq-hg}. It thus requires higher brightness scaling factors for larger $g$ to reproduce the nearly face-on disk for J1604. On the other hand, $\alpha_{\rm in}$ and $\alpha_{\rm our}$ describes the asymptotic radial power law indices interior and exterior to $R_c$. To reproduce any disk surface brightness distribution, an increase in $R_c$ requires the decrease in the positive $\alpha_{\rm in}$ (or a decrease in the negative $\alpha_{\rm our}$) to produce the surface brightness interior to (or exterior to) it in Eq.~\eqref{eq-disk}, since otherwise the interior regions would be too faint (or the exterior regions would be too bright). %, prompting us to attempt deriving them based on a few straightforward assumptions. %Assuming certain mathematical identities, we can derive relationships among various parameters and then ploted it on the corner figures.

% \begin{figure*}[htb!]
%   \centering
%   \includegraphics[width=1.0\textwidth]{K1-corner.pdf}
%   \caption{\textcolor{red}{MCMC modeling results using forward modeling for the $K1$ band.}} \label{fig:fig-K1-corner}
% \end{figure*}

% \begin{figure*}[htb!]
%   \centering
%   \includegraphics[width=1.0\textwidth]{K2-corner.pdf}
%   \caption{\textcolor{red}{MCMC modeling results using forward modeling for the $K2$ band.}} \label{fig:fig-K2-corner}
% \end{figure*}

% \begin{figure*}[htb!]
%   \centering
%   \includegraphics[width=1.0\textwidth]{L-corner.pdf}
%   \caption{\textcolor{red}{MCMC modeling results using forward modeling for the $L'$ band.}} \label{fig:fig-L-corner}
% \end{figure*}

\end{appendix}

\end{CJK*}
\end{document}